\documentclass[twocolumn,aps,prl,longbibliography]{revtex4-2}

\usepackage{amsmath}
\usepackage{amsfonts}
\usepackage{amssymb}
\usepackage{amsthm}
\usepackage{graphicx}
\usepackage{bbm}
\usepackage{hyperref}
\usepackage{xcolor}

\newtheorem{SPTO_Contextuality}{Lemma}
\newtheorem{SPTO_Games}{Theorem}
\newtheorem{SPTO_Robust}[SPTO_Games]{Theorem}
\newtheorem{SPTO_Advantage}{Corollary}

\usepackage{enumitem}

\begin{document}

\title{Quantum computational advantage with string order parameters of 1D symmetry-protected topological order}

\author{Austin K. Daniel}
\email{austindaniel@unm.edu}
\author{Akimasa Miyake}
\email{amiyake@unm.edu}
\affiliation{Center for Quantum Information and Control, Department of Physics and Astronomy, University of New Mexico, Albuquerque, NM 87131, USA}

\begin{abstract}
Nonlocal games with advantageous quantum strategies give arguably the most fundamental demonstration of the power of quantum resources over their classical counterparts.  Recently, certain multiplayer generalizations of nonlocal games have been used to prove unconditional separations between limited computational complexity classes of shallow-depth circuits. Here, we show advantageous strategies for these nonlocal games for generic ground states of one-dimensional symmetry-protected topological orders (SPTO), when a discrete invariant of SPTO known as a twist phase is nontrivial and -1.  Our construction demonstrates that sufficiently large string order parameters of such SPTO are indicative of globally constrained correlations useful for the unconditional computational separation. 
\end{abstract}

\date{\today}
\maketitle

{\textit{Introduction.---}}
Entanglement underlies nonclassical features of quantum mechanics.  On one hand, 
local hidden variable models cannot produce nonlocal quantum correlations \cite{einstein1935can, bell1964on}.  This idea is elegantly illustrated with nonlocal games \cite{cleve2004consequences, brunner2014bell}, whereby players who implement strategies utilizing entangled resources can accomplish a distributed computational task without classical communication. Moreover, in Ref.~\cite{barrett2007modeling}, it was shown that local hidden variable models assisted by even a limited amount of classical communication fail to mimic Pauli-measurement outcomes on graph states \cite{hein2005entanglement}. On the other hand, contextuality \cite{kochen1967the, peres1991two, cabello2010non, abramsky2011the, cabello2014graph}, the degree to which locally incompatible measurements evade global explanation, is another nonclassical feature related to the hardness of computation and quantum advantage \cite{ raussendorf2013contextuality, howard2014contextuality, abramsky2017contextual, bermejo2017contextuality, raussendorf2017contextuality, frembs2018contextuality, mansfield2018quantum, beer2018contextuality}. Combining these features, seminal works by Bravyi \textit{et. al.} \cite{bravyi2018quantum} and others \cite{coudron2018trading, gall2018average, bravyi2020quantum, watts2019exponential, grier2019interactive} compared certain many-body generalizations of nonlocal games assisted by limited classical communication to classical computation with bounded fan-in gates. This perspective is successful in proving unconditional exponential separations between limited computational complexity classes, demonstrating the power of shallow quantum circuits over their classical counterparts.

Advantageous quantum strategies for these multiplayer games  
possess two key properties; contextuality of the measurements performed and long-range entanglement accessible by arbitrarily distant players.  Motivated by this key observation, we establish a general connection between the shared quantum resource and many-body entanglement ubiquitously present in ground states of quantum phases of matter called symmetry-protected topological order (SPTO) \cite{pollmann2010entanglement, pollmann2012symmetry, chen2013symmetry, zeng2019quantum}.
Namely, we show that local measurements that collectively resolve global measurements of symmetries and so-called twist phases \cite{hung2014universal} (an invariant of 1D SPTO phases with an abelian symmetry group) give a desired state-dependent contextuality property.  Furthermore, the string order parameter \cite{nijs1989preroughening, perez2008string}, a nonlocal order parameter of 1D SPTO related to the long-ranged order \cite{verstraete2004entanglement, verstraete2004diverging, popp2005localizable, venuti2005analytic, symmetry2017marvian}, gives the desired entanglement structure, which is known to be useful for measurement-based quantum computation (MBQC) \cite{miyake2010quantum, bartlett2010quantum, else2012symmetry, else2012symmetry2, miller2015resource, raussendorf2017symmetry, stephen2017computational, devakul2018universal, raussendorf2019computationally, stephen2019subsystem, daniel2020computational}.

Our work indicates that the aforementioned computational separation between shallow-depth classical and quantum circuits carries over to generic 1D SPTO ground states. 
 This will be illustrated using various states in the 1D $\mathbb{Z}_2\times\mathbb{Z}_2$ SPTO phase, such as the cluster state \cite{briegel2001persistent, raussendorf2001a} and the Affleck-Kennedy-Tasaki-Lieb (AKLT) state \cite{affleck1987rigorous}.  It is intriguing to see how the string-like correlations of 1D SPTO states have similar utility as the two-point correlations of the Greenberger-Horne-Zeilinger (GHZ) state, as the so-called GHZ paradox \cite{mermin1990simple} has been a canonical example in nonlocal games and nonadaptive MBQC \cite{anders2009computational, hoban2011non-adaptive, hoban2011generalized, raussendorf2013contextuality, hoban2014measurement}.  Our result assists to tighten an inherent connection between MBQC, contextuality, and group cohomology pursued in Refs.~\cite{abramsky2012the, abramsky2015contextuality, okay2017topological, okay2018cohomological, raussendorf2019cohomological, aasnaess2020cohomology}. In comparison, however, our obstruction to a noncontextual description of the triangle game below arises directly from a cohomological signature of 1D SPTO.  Our results also complement studies of nonlocality in many-body systems \cite{deng2012bell, tura2014detecting, tura2017energy, chiara2018genuine}.   Last but not least, as quantum simulation of various 1D SPTO states is of broad interest in experimental realizations \cite{endres2011observation, senko2015realization, deleseleuc2019observation}, our construction may pave a way towards observation of quantum computational advantage using 1D SPTO and its string order parameter.

\begin{figure*}
\includegraphics[width=\linewidth]{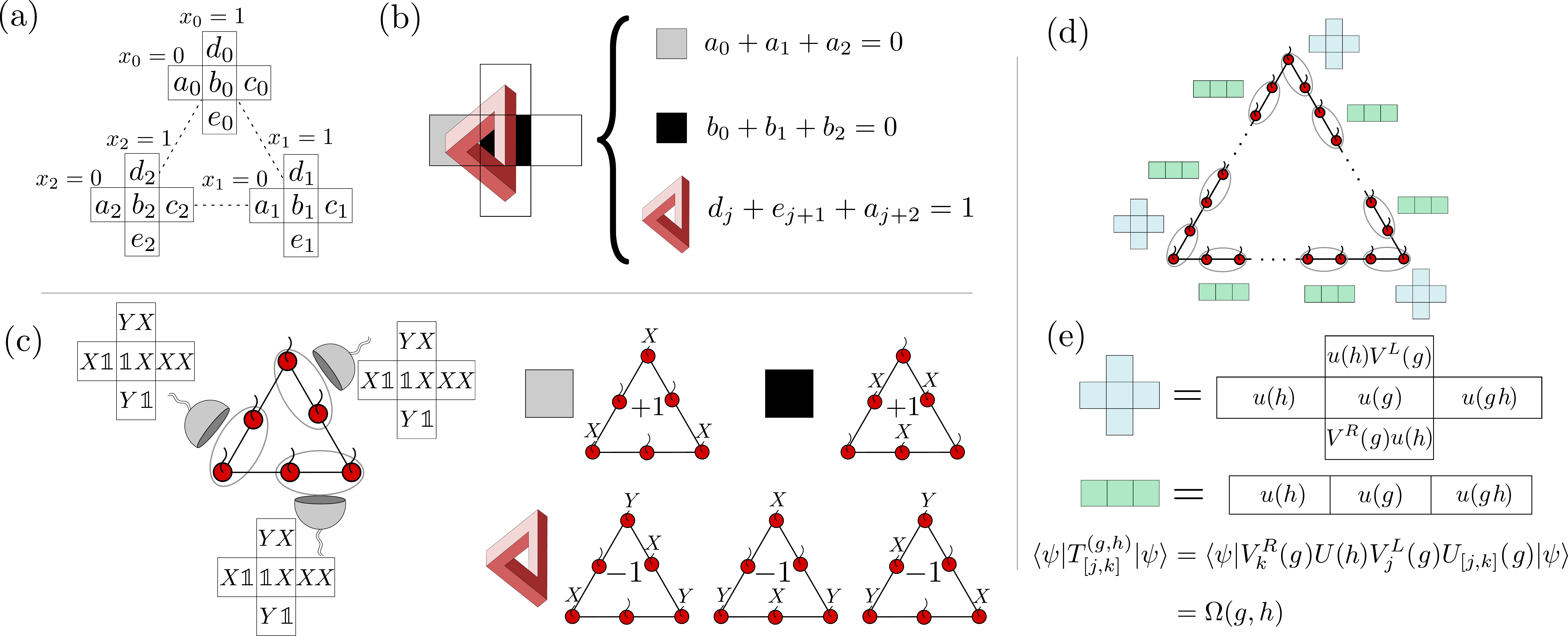}
\caption{(Color online) Triangle game. (a) Players fill in the row or column of their table with a binary string if their input is 0 or 1, respectively. (b) The win conditions. Apart from the global even parity of the output, the dark and light shaded boxes denote that each entry in the row jointly have even parity.  The Penrose triangle represents the condition that the top, bottom, and left entries of any clockwise ordering of the three players have odd parity. (c) Quantum strategy for the triangle game. For each pair of Pauli observables in the table, the left one corresponds to the qubit located at the corresponding corner of the triangle. Perfection of the strategy is ensured by five cluster state stabilizers, whose eigenvalues are $\pm1$ as shown. (d) Multiplayer triangle game (see \cite{SM}). Three arbitrary players, depicted at the corners of the triangle, measure the same Pauli observables as before on the $2n$-qubit cluster state and otherwise measure along the row. They still win the original game perfectly (up to inconsequential additional outputs by the other $n-3$ players). (e) Perfect quantum strategy on 1D SPTO fixed-point states is ensured when players measure on-site symmetry and boundary operators.  The Penrose-triangle constraints in (c) manifest as a collective measurement of twisted string order parameters whose expectation value is an invariant of SPTO, called a twist phase $\Omega(g,h)$, equal to $-1$.}
\label{Triangle_Game}
\end{figure*}

{\textit{The triangle game.---}} We begin with a motivating example adapted from Refs.~\cite{barrett2007modeling, bravyi2018quantum}, as seen in Fig.~\ref{Triangle_Game}.
Consider a game where three players, indexed by $j\in\{0,1,2\}$, each receive a random input bit $x_j\in\{0,1\}$. Each fills a three-bit string $\mathbf{y}_j \in\{0,1\}^3$ in the row or column of the table of Fig.~\ref{Triangle_Game}(a) if $x_j =0$ or $1$, respectively.  Suppose the players do not communicate and produce outputs dependent only on their given input, i.e. $\mathbf{y}_j = \mathbf{y}_j(x_j)$. The values recorded in the row or column of each table can be written $\mathbf{y}_j(0) = (a_j, b_j, c_j)$ and $\mathbf{y}_j(1) = (d_j, b_j, e_j)$.  The players win the game whenever the full output string $(\mathbf{y}_0,\mathbf{y}_1,\mathbf{y}_2)\in\{0,1\}^9$ has even parity and
\begin{subequations}
\begin{align}
a_0 + a_1 + a_2 &=0 \label{Tri_Win_1} ,\\
b_0 + b_1 + b_2 &=0 \label{Tri_Win_2} ,\\
d_{j} + e_{j+1} +a_{j+2} &= 1~~\forall j\in\{0,1,2\} \label{Tri_Win_3}.
\end{align}
\end{subequations}
 Notice that while Eq.~(\ref{Tri_Win_2}) must hold for all inputs, Eqs.~(\ref{Tri_Win_1}) and (\ref{Tri_Win_3}) are input dependent constraints. However, because summing Eqs.~(\ref{Tri_Win_1})-(\ref{Tri_Win_3}) gives $\sum_{j=0}^2 (d_j + b_j + e_j) =1$, the total output string for the input $\mathbf{x}=(1,1,1)$ cannot have even parity. This implies that the classical winning probability is bounded above by $\frac{7}{8}$, by failing on at least one of eight inputs.

On the contrary, there is a perfect quantum strategy for this game. A quantum strategy for a nonlocal game is a tuple $(\rho,\mathcal{C}_x)$ consisting of a shared quantum state $\rho$ and \textit{contexts}, sets of pairwise commuting local observables $\mathcal{C}_x=\{A_k(x)\}_k$ to be measured, for each $x\in\{0,1\}$.  Let $X$, $Y$, and $Z$ be the Pauli matrices and $\mathbbm{1}$ be the identity matrix.  Each player $j$ holds qubits $2j$ and $2j+1$ from the six-qubit 1D cluster state, $|\psi_{\textrm{1DC}}\rangle = \prod_{k=0}^5 CZ_{k,k+1}|+\rangle^{\otimes 6}$, where $|+\rangle = (|0\rangle + |1\rangle)/\sqrt{2}$ and $CZ = |0\rangle\langle 0 | \otimes \mathbbm{1} + |1\rangle\langle 1|\otimes Z$ is the two-qubit controlled-$Z$ gate. $|\psi_{\textrm{1DC}}\rangle$ is a stabilizer state, the joint +1 eigenstate of a commuting set of Pauli observables generated by $Z_k X_{k+1} Z_{k+2}$ for $k = 0,...,5$.  Each player measures the two-qubit Pauli observables from the horizontal or vertical contexts shown in Fig.~\ref{Triangle_Game}(c) and records the outcomes in the table. The observables in each row and column multiply to the identity, constraining the measurement outcomes to form a string of even parity.  Certain observables in each player's table collectively form stabilizers up to a sign, shown in Fig.~\ref{Triangle_Game}(c), implying Eqs.(\ref{Tri_Win_1})-(\ref{Tri_Win_3}) are satisfied.  These stabilizers form an identity product \cite{waegell2019benchmarks}, giving state-dependent contextuality.

Moreover, as described in Refs.~\cite{barrett2007modeling, bravyi2018quantum}, this game has a multiplayer generalization (see \cite{SM} for details).  In each round of the game three arbitrary players, labeled $\alpha$, $\beta$, and $\gamma$, are given a bit from the input $\mathbf{x}\in\{0,1\}^3$ and each player outputs a three-bit string.  The new win conditions are equivalent to Eqs.~(\ref{Tri_Win_1})-(\ref{Tri_Win_3}) up to the parity of a ``correction string" given by the outputs of the other players.  The corresponding quantum strategy utilizes a $2n$-qubit 1D cluster state and measurements in the same contexts, as depicted in Fig.~\ref{Triangle_Game}(d).  In this generalized $n$-player setting, quantum players can outperform even locally \emph{communicating} classical players.  For large enough $n$, a constant number of rounds of classical communication between players that are nearby with respect to the cycle cannot create the same global correlations attained in the quantum setting since distant players cannot communicate. 
These nonclassical, long-ranged correlations that persist for the quantum strategy in the large-$n$ (i.e. thermodynamic) limit are naturally indicative of 1D symmetry-protected topological order.

{\textit{Symmetry-protected topological order (SPTO).---}}
A 1D SPTO phase is topologically ordered in the presence of symmetry $G$, in that each ground state cannot be connected smoothly to a product state via symmetry-respecting perturbations. In the following, we focus on a global symmetry $G$ that forms a finite group. The topological nature gives ground-state degeneracy dependent on boundary conditions.  At the open boundary, there appear effective degrees of freedom that transform under a \emph{projective representation} of $G$.

Algebraically, a projective representation of $G$ is a collection of unitary matrices $\{V(g)\}_{g\in G}$ that obey the group multiplication law up to a $G$-dependent phase, i.e.,
\begin{align}
V(g)V(h) = \omega(g,h)V(gh),
\end{align}
where $\omega(g,h)\in U(1)$ is called a 2-cocycle.  
Inequivalent projective representations, and thus 1D SPTO phases, are classified by a multiplicative group $H^2(G,U(1))$ called the second group cohomology \cite{chen2011classification}, whose elements are equivalence classes of 2-cocycles, called cohomology classes, denoted $[\omega]\in H^2(G,U(1))$.

{\textit{Symmetry twists.---}}
The cohomological properties can be probed, even under periodic boundary conditions, by introducing artificial boundaries called symmetry twists \cite{hung2014universal, zaletel2014detecting}.  Consider a system of $n$ sites with global symmetry $G$ carrying on-site representation $U(g)= u(g)^{\otimes n}$. Denote as $U_{[j,k]}(g) = \otimes_{j+1}^{k-1}u(g)$ a truncated symmetry operator acting only between sites $j$ and $k$. Symmetry twists are low energy excitations that appear about sites $j$ and $k$ when $U_{[j,k]}(g)$ acts on the 1D SPTO ground state $|\psi\rangle$. 
 In general, there are local operators $V_j^L(g)$ and $V_k^R(g)$, called \textit{boundary operators}, supported in the vicinity of sites $j$ and $k$ that annihilate the symmetry twists.  
Mathematically, this is realized by the trivial action of
\begin{align}
S_{[j,k]}(g) = \left(V^L_j(g)\otimes V^R_k(g)\right)U_{[j,k]}(g)
\label{Trunc_Sym}
\end{align}
on the state, i.e., $S_{[j,k]}(g) |\psi\rangle = |\psi\rangle$.  The expectation value of $S_{[j,k]}(g)$ gives a string order parameter that characterizes the long-range order in the 1D SPTO phase \cite{nijs1989preroughening, perez2008string}.
For translationally invariant systems, one may drop the site dependence on $V^L_j(g)$ and $V^R_k(g)$.

We remark that the boundary operators are not universal (i.e., they vary for different states in the phase).  Thus we focus on fixed-point boundary operators, which are defined with respect to the fixed-point state of the 1D SPTO phase obtained under renormalization group flow \cite{verstraete2005renormalization, huang2013symmetry, zeng2019quantum}.  Hereafter, we redefine $V^L(g)$ and $V^R(g)$ to be the fixed-point boundary operators,  which have local support of size $m$, typically two. In \cite{SM}, we construct $V^L(g)$ and $V^R(g)$ explicitly from matrix-product state (MPS) representations of the fixed-point state.
In the $[\omega]$-class 1D SPTO phase, operators $\{V^R(g)\}_{g\in G}$ and $\{V^L(g)\}_{g\in G}$ form projective representations of $G$ residing in cohomology class $[\omega]$ and $[\omega^*]$, respectively.  At the same site, they satisfy 
\begin{align}
V^R(g)V^R(h) &= \omega(g,h) V^R(gh) \label{R_Proj} ,\\ 
V^L(g)V^L(h) &= \omega(g,h)^* V^L(gh) \label{L_Proj} ,\\
V^R(g)V^L(h) &= V^L(h)V^R(g) \label{RL_Com} ,\\
V^R(g)V^L(g)&= u(g)^{\otimes m} \label{RL_Sym}.
\end{align}
We prove Eqs.~(\ref{R_Proj})-(\ref{RL_Sym}) in \cite{SM}.

{\textit{Twist phase and twisted string order parameter.---}}
1D SPTO phases possess an invariant called a twist phase $\Omega(g,h)\in U(1)$ \cite{hung2014universal} defined as,
\begin{align}
\Omega(g,h) = \frac{\omega(g,h)}{\omega(h,g)}.
\label{Twist_Def}
\end{align}
For abelian $G$, this object depends only on the cohomology class $[\omega]$. 
Conveniently, this phase is simply the overall phase accumulated upon commuting the projective representations of $g$ and $h$ through each other, i.e. $V(g)V(h) = \Omega(g,h)V(h)V(g)$. 

In comparison to Eq.~(\ref{Trunc_Sym}), it is convenient to define the ``twisted'' string order parameter as the expectation value of an operator
\begin{align}
T_{[j,k]}^{(g,h)} &= V_{k}^R(g) U(h) V_{j}^L(g) U_{[j,k]}(g).
\label{Twist_Op}
\end{align}
By Eqs.~(\ref{Trunc_Sym})-(\ref{RL_Sym}), its expectation value on the fixed-point state is the twist phase,
\begin{align}
\langle\psi|T_{[j,k]}^{(g,h)} |\psi\rangle = \Omega(g,h).
\label{Twist_Phase_Expectation}
\end{align}
See \cite{SM} for a proof of Eq.~(\ref{Twist_Phase_Expectation}).

{\textit{SPTO triangle game strategy from symmetry twists.---}}
Now we present the main result of of this paper.  We show that the measurement of twist phases for a particular class of 1D SPTO phases can be repurposed as a quantum strategy for the multiplayer triangle game.  As onsite symmetries will always be measured over $m$ sites,  henceforth we redefine $u(g)$ to denote $u(g)^{\otimes m}$ for ease of the notation.

\begin{SPTO_Contextuality}
Consider a 1D SPTO ground state with a finite abelian symmetry group $G$ containing elements $g,h\in G$ such that the twist phase $\Omega(g,h)=-1$. There are two overlapping contexts of local observables by which the twisted string order parameter $T_{[j,k]}^{(g,h)}$ of Eq.~(\ref{Twist_Op}) is composable.
\label{SPTO_Context}
\end{SPTO_Contextuality}

\textit{Proof:}  The operators appearing in $T_{[j,k]}^{(g,h)}$ can be organized in the following table,
\begin{align}
\vcenter{\hbox{\includegraphics[width=0.55\linewidth]{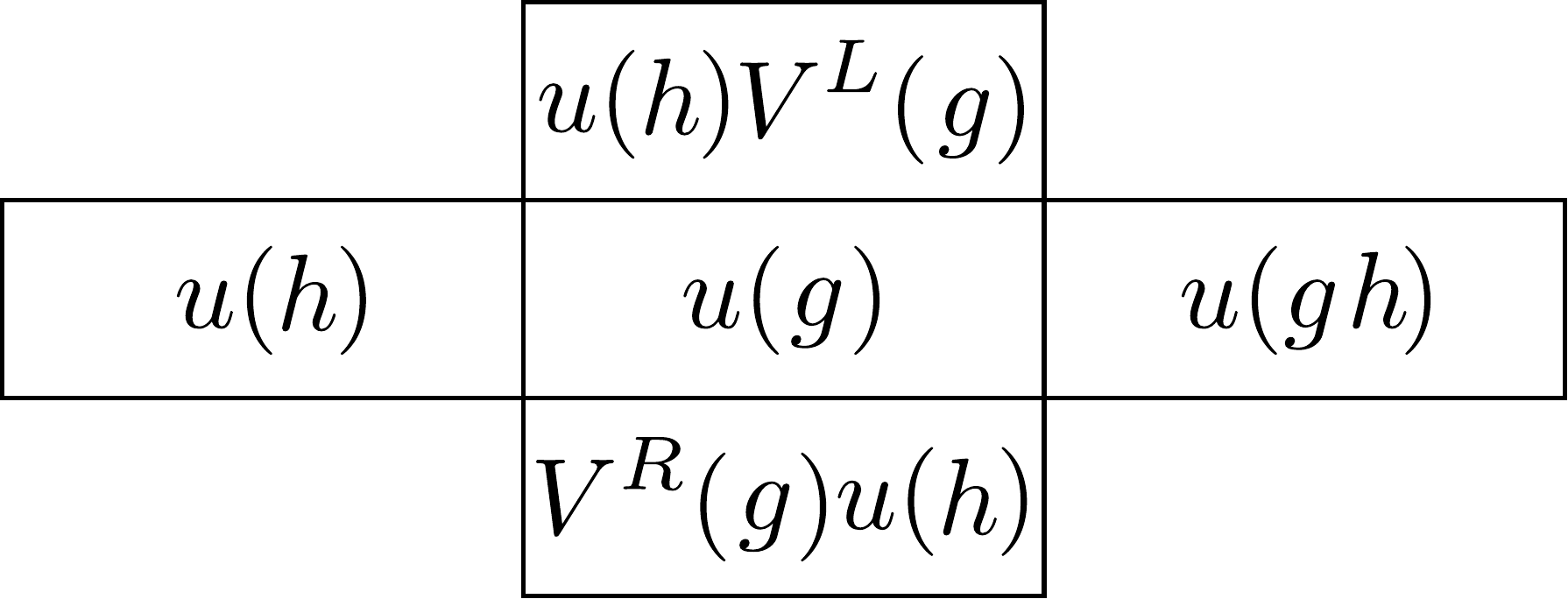}}}~.
\label{Tri_Sqr}
\end{align}
Since $G$ is abelian, all on-site symmetry operators in the row commute. By Eq.~(\ref{RL_Com}) and (\ref{RL_Sym}), $u(g)$ commutes with $u(h)V^L(g)$ and $V^R(g)u(h)$.   Finally, by Eqs.~(\ref{R_Proj})-(\ref{RL_Com}), $\left( u(h)V^L(g) \right)\left( V^R(g)u(h) \right) = \Omega(g,h)^2 \left( V^R(g)u(h) \right)\left( u(h)V^L(g) \right)$.
Thus the operators in the column commute if and only if $\Omega(g,h) = \pm1$.~~$\square$

\begin{SPTO_Games}
Consider a 1D SPTO phase with a symmetry group $G$ as described in Lemma~\ref{SPTO_Context}. Any fixed-point ground state in the phase allows a perfect quantum strategy for the multiplayer triangle game.
\label{SPTO_Games}
\end{SPTO_Games}
\textit{Proof:} Suppose each player holds a block of $m$ constituent particles from the 1D SPTO fixed-point $|\psi\rangle$.  Each player measures their block in the horizontal or vertical context of Eq.~(\ref{Tri_Sqr}) if they are given input 0 or 1, respectively.  By Eq.~(\ref{RL_Sym}), the product of all operators in either context of Eq.~(\ref{Tri_Sqr}) is $u(g^2h^2)$, so   
collectively the players measure global symmetry $U(g^2h^2)$ and the product of all outcomes is $+1$.  Regardless of the input, each player measures $u(g)$ and thus they collectively measure global symmetry $U(g)$, implying Eq.~(\ref{Tri_Win_2}) is satisfied. For the input $(0,0,0)$, each player measures $u(h)$, so collectively they measure global symmetry $U(h)$, implying Eq.~(\ref{Tri_Win_1}) is satisfied. Finally, for the input $(1,1,0)$, player 0 measures $u(h)V^L(g)$, player 1 measures $V^R(g)u(h)$, and player 2 measures $u(h)$. Collectively they measure the three-site twisted string order parameter $T^{(g,h)}_{[0,1]}$ and the joint outcome is $\Omega(g,h)=-1$, by Eq.~(\ref{Twist_Phase_Expectation}). Permutations of this argument show that Eq.~(\ref{Tri_Win_3}) is satisfied. The strategy for the multiplayer version follows accordingly.~~$\square$

{\textit{Examples in the }$\mathbb{Z}_2\times\mathbb{Z}_2$\textit{ SPTO phase.---}}
The simplest SPTO phase in which Thm.~\ref{SPTO_Games} holds is the nontrivial $\mathbb{Z}_2\times\mathbb{Z}_2$ 1D SPTO phase.
The complete set of twist phases, given by $\Omega((a,b),(c,d))= (-1)^{ad+bc}$ for $(a,b),(c,d)\in\mathbb{Z}_2\times\mathbb{Z}_2$, is identical to the Pauli algebra.
We show how Thm.~\ref{SPTO_Games} encompasses the quantum strategy for the triangle game discussed above, and then extend Thm.~\ref{SPTO_Games} to generic states outside the fixed-point.
 We illustrate these results using the 1D cluster and AKLT states, respectively. Both are known to be useful as 1D quantum logical wires in MBQC.  \cite{raussendorf2001a, raussendorf2003measurement, gross2007novel, gross2007measurement, brennen2008measurement}
 
The 1D cluster state \cite{briegel2001persistent, raussendorf2001a} is the fixed-point of this phase. The on-site symmetry and boundary operators are 
\begin{subequations}
\begin{align}
u((a,b)) &= X^a\otimes X^b ,\label{Sym_1DC}\\
V^R((a,b)) &= Z^bX^a\otimes Z^a ,\label{VR_1DC}\\
V^L((a,b)) &= Z^b\otimes Z^aX^b \label{VL_1DC},
\end{align}
\end{subequations}
for $(a,b)\in\mathbb{Z}_2\times\mathbb{Z}_2$.  Taking $g=(0,1)$ and $h=(1,0)$ in Eq.~(\ref{Tri_Sqr}) gives the strategy presented in Fig.~\ref{Triangle_Game}(c).

To study generic states beyond the fixed-point, we will refer the set of measurements to be performed as the protocol.  The protocol corresponding to the contexts of Eq.~(\ref{Tri_Sqr}) constructed with the fixed-point boundary operators will be referred to as the \textit{fixed-point protocol}.  The quantum strategy formed by the fixed-point protocol implemented on arbitrary states in the phase no longer wins with unit probability, but extends Theorem.~\ref{SPTO_Games} as follows.

\begin{SPTO_Robust}
Consider an arbitrary 1D $\mathbb{Z}_2\times\mathbb{Z}_2$ SPTO ground state $|\phi\rangle$ and let $\langle\mathcal{S}\rangle = \mathrm{min}_{g\in G}\{\langle\phi| S_{[j,k]}(g)|\phi\rangle\}$ be the minimal value of any string order parameter constructed from the fixed-point boundary operators.  The fixed-point protocol of Theorem~\ref{SPTO_Games} implemented on the state $|\phi\rangle$ yields an advantageous quantum strategy (i.e. $\mathrm{pr}(\mathrm{win})>7/8$) whenever $1/3<\langle\mathcal{S}\rangle\leq1$.
\label{SPTO_Robust}
\end{SPTO_Robust}
\textit{Proof:}  In the $\mathbb{Z}_2\times\mathbb{Z}_2$ SPTO phase $S_{[j,k]}(g)^2=1$ and each ground state $|\phi\rangle$ is symmetric.  Denote by $\textrm{pr}_{\phi}(\pm1|O)$ the probability that the joint measurement outcome of a dichotomic observable $O$ on $|\phi\rangle$ has parity $\pm1$.  By definition, $\textrm{pr}_{\phi}(\pm1|O) = (1\pm\langle\phi| O |\phi\rangle)/2$, so $\textrm{pr}_{\phi}(+1|U(g))=1~\forall g$. Because $T^{(g,h)}_{[j,k]} = \Omega(g,h) U(h)S_{[j,k]}(g)$ by Eq.~(\ref{RL_Sym}), $\textrm{pr}_{\phi}(-1|T^{(g,h)}_{[j,k]})=\textrm{pr}_{\phi}(+1|S_{[j,k]}(g))\geq(1+\langle\mathcal{S}\rangle)/2$ when  $\Omega(g,h)=-1$.  Averaging over the eight possible inputs for the triangle game, we find $\textrm{pr}(\textrm{win}) = \frac{5}{8} \textrm{pr}_{\phi}(+1|U(g)) + \frac{3}{8} \textrm{pr}_{\phi}(-1|T^{(g,h)}_{[j,k]}) \geq (13 + 3\langle \mathcal{S}\rangle)/16$.  Thus, $\textrm{pr}(\textrm{win})>7/8$ whenever $1/3<\langle\mathcal{S}\rangle\leq1$.~~$\square$

Thm.~\ref{SPTO_Robust} extends the quantum advantage to more realistic states that are neither fixed-point states nor Pauli stabilizer states. The AKLT state \cite{affleck1987rigorous} is a spin-1 antiferromagnetic state that is the $\mathbb{Z}_2\times\mathbb{Z}_2$ SPTO ground state of a two-body interacting Hamiltonian (as opposed to the three-body interactions of the 1D cluster state).  Over two spin-1's, we introduce $|\tilde{e}\rangle$ as the singlet, and $\{|\tilde{x}\rangle,|\tilde{y}\rangle,|\tilde{z}\rangle\}$ as a Cartesian basis of the triplet (see \cite{SM} for mathematical definitions).  Using $\mu\in\{z,x,y\}$ to denote group elements $\{(0,1),(1,0),(1,1)\}$, respectively, the on-site symmetry and fixed-point boundary operators are
\begin{subequations}
\begin{align}
u(\mu) &= \exp\left(i\pi S^\mu\right)^{\otimes 2} ,\label{Sym_AKLT}\\
V^R(\mu) &= |\tilde{e}\rangle\langle \tilde{\mu}| + |\tilde{\mu} \rangle\langle \tilde{e}| + i\sum_{\nu,\gamma\in\{x,y,z\}} \epsilon_{\mu\nu\gamma} |\tilde{\nu}\rangle\langle \tilde{\gamma}| , \label{VR_AKLT}\\
V^L(\mu) &= |\tilde{e}\rangle\langle \tilde{\mu}| + |\tilde{\mu} \rangle\langle \tilde{e}| - i\sum_{\nu,\gamma\in\{x,y,z\}} \epsilon_{\mu\nu\gamma} |\tilde{\nu}\rangle\langle \tilde{\gamma}| \label{VL_AKLT}.
\end{align}
\end{subequations}
The string order parameter formed by these operators (as per Eq.~(\ref{Trunc_Sym})) consists of dichotomic operators, in contrast to the conventional one based on spin-1 operators \cite{nijs1989preroughening}. 
Note however that Eqs.~(\ref{Sym_AKLT})-(\ref{VL_AKLT}) are equivalent to Eqs.~(\ref{Sym_1DC})-(\ref{VL_1DC}) under a local isometry $|\tilde{e}\rangle\langle ++| + |\tilde{z}\rangle\langle -+| + |\tilde{x}\rangle\langle +-| + i|\tilde{y}\rangle\langle--|$, as the fixed-point is the 1D cluster state.
For $g=z$ and $h=x$, an exact MPS calculation
gives $\langle\mathcal{S}\rangle \geq \frac{4}{9}\left( \sqrt{\frac{2}{3}} + \frac{2}{3}\right)^2 \approx 0.978$ and by Thm.~\ref{SPTO_Robust}, $\textrm{pr}(\textrm{win}) \geq \frac{13}{16} + \frac{1}{12}\left( \sqrt{\frac{2}{3}} + \frac{2}{3}\right)^2 \approx 0.996$  (See \cite{SM} for details).  Thus quantum advantage persists at the AKLT point.

\textit{Quantum computational advantage.---} 
In Ref.~\cite{bravyi2018quantum}, an exponential quantum speed-up was shown for a problem equivalent to a 2D multiplayer generalization of our triangle problem where players are situated on an $N\times N$ grid (elaborated in \cite{SM}).  In this 2D setting, quantum players outperform nonlocally communicating classical players.  Indeed, a constant number of rounds of classical communication between a constant number of arbitrarily distant players on the grid still leaves at least one cycle of locally communicating players
in the large-$N$ limit.  This advantage can be rephrased in the language of circuit complexity. Classical Boolean circuits consisting of nonlocal gates with bounded fan-in require at least logarithmic depth (i.e. logarithmically many rounds of communication) to ensure a solution to the problem with arbitrarily high probability.
On the other hand, it is possible to prepare generic SPTO ground states in constant depth when the symmetry $G$ is disregarded \cite{chen2010local,hastings2005quasiadiabatic}. 
Thms.~\ref{SPTO_Games} and \ref{SPTO_Robust} present a substantial extension regarding the required capability of a quantum device.
\begin{SPTO_Advantage}
Consider a relation problem whereby players situated on an $N\times N$ 2D grid are tasked to play the multiplayer triangle game on an arbitrary cycle in the grid.  A quantum device that can prepare a 1D SPTO ground state in constant time with string order parameters greater than 1/3 on the arbitrary cycles and perform the fixed-point protocol of Theorem~\ref{SPTO_Robust} solves the problem with probability greater than 7/8 on all inputs, which any classical circuit with gates of fan-in at most $K$ and depth less than $ \log_K(N)$ cannot do.
\label{SPTO_Advantage}
\end{SPTO_Advantage}
\noindent A precise statement and proof of Cor.~\ref{SPTO_Advantage} is given in \cite{SM}.

{\textit{Conclusion and outlook.---}}
We have shown how to harness contextuality and the string order parameter of generic 1D SPTO ground states to construct advantageous quantum strategies for a nonlocal game that thwarts all classical strategies (even with assistance of limited long-range communication).  Our approach, to be supplemented with a follow-up paper \cite{daniel2020prep}, contributes to unify recent insight about unconditional quantum advantage.  For example, the magic-square game in \cite{bravyi2020quantum} also admits general 1D SPTO strategies, and similar complexity-theoretic results using the GHZ state \cite{watts2019exponential} can be understood using Kennedy-Tasaki duality maps \cite{kennedy1992hidden, else2013hidden}.  Our relation of the string order parameter to robustness of the advantage may be applicable to robust self-testing \cite{mayers2004self, natarajan2017a, coladangelo2017robust, supic2019self, baccari2020device} for fixed-point SPTO states.  The use of SPTO is welcome in scalable and robust experimental demonstrations, as these ground states can be realized as the unique ground states of two-body Hamiltonians in contrast to the GHZ state. Broadly, our work is timely to promote the value of quantum simulation to prepare and detect 1D SPTO for potential quantum advantage.

\begin{acknowledgments}
{\textit{Acknowledgments.---}}
We thank V. Buchemmavari and A. Zhao for helpful discussions.  This work was supported by NSF grants PHY-1620651, PHY-1818914, and PHY-1915011, and the Department of Energy Center Quantum Systems Accelerator.
\end{acknowledgments}

\bibliography{ref.bib}

\begin{widetext}

\end{widetext}

\end{document}


\title{Supplemental Material for ``Quantum computational advantage with string order parameters of 1D symmetry-protected topological order"}

\author{Austin K. Daniel}
\email{austindaniel@unm.edu}
\author{Akimasa Miyake}
\email{amiyake@unm.edu}
\affiliation{Center for Quantum Information and Control, Department of Physics and Astronomy, University of New Mexico, Albuquerque, NM 87131, USA}

\begin{abstract}
This supplemental material is comprised of three main sections.  In Sec.~\ref{SM1} we discuss the multiplayer triangle game and formally state and prove Cor.~\ref{SPTO_Advantage} from the text.  In Sec.~\ref{SM2} we construct the fixed-point boundary operators used in the text from MPS representations of the fixed-point of the 1D SPTO phase.  Finally, in Sec.~\ref{SM3} we use the MPS representation of the AKLT state to calculate the success probability of quantum strategy for the multiplayer triangle game from the fixed-point protocol implemented at the AKLT point.
\end{abstract}

\date{\today}
\maketitle

\section{The multiplayer triangle game and formal statement of Corollary~1}
\label{SM1}

\subsection{Multiplayer triangle game}
\label{N_Player}
In this section we formally state the win conditions for the multiplayer triangle game.  Consider a scenario where $n$ players located around a cycle are enumerated as players 0 through $n-1$ in a clockwise manner.  Suppose that three players with labels $\alpha$, $\beta$, and $\gamma$ with $\alpha<\beta<\gamma$ are given one bit each from a three-bit input string $\mathbf{x} = (x_\alpha, x_\beta, x_\gamma)\in\{0,1\}^3$ drawn uniformly at random.   Each player is tasked to fill in the row or column of their respective table with a three-bit output string $\mathbf{y}_j$ depending on if their received input is 0 or 1, respectively.  All players other than $\alpha$, $\beta$, and $\gamma$ receive input 0. We may depict players $\alpha$, $\beta$, and $\gamma$ as residing at the top, lower right, and lower left corners of a triangle, respectively.  If we denote the output bit string as a function of the input, i.e. $\mathbf{y}_j = \mathbf{y}_j(x_j)$, we can write $\mathbf{y}_j(0) = (a_j, b_j, c_j)$ and $\mathbf{y}_j(1) = (d_j, b_j, e_j)$ in correspondence with the table.  The players are said to win the game whenever the joint output $\{\mathbf{y}_j\}_{j=0}^{n-1}$ forms a string of even parity (i.e. $\sum_{j=0}^{n-1} |\mathbf{y}_j|=0 \textrm{ mod }2$, where $|\cdot|$ denotes the Hamming weight) and the following equations hold
\begin{subequations}
\begin{align}
a_\alpha + a_\beta + a_\gamma &= a_R + a_L + a_B \label{N_Tri_1}\\
b_\alpha + b_\beta + b_\gamma &= b_R + b_L + b_B \label{N_Tri_2}\\
d_\alpha + e_\beta + a_\gamma &= 1+c_R + a_B + a_L\label{N_Tri_3}\\
d_\beta + e_\gamma + a_\alpha &= 1+c_B + a_L + a_R \label{N_Tri_4}\\
d_\gamma + e_\alpha + a_\beta &= 1+c_L + a_R + a_B \label{N_Tri_5},
\end{align}
\end{subequations}
where using joint indices for the right (R) edge, bottom (B) edge, and left (L) edge of the triangle, we define for $\sigma\in\{a,b,c\}$,
\begin{subequations}
\begin{align}
\sigma_R &= \sum_{j=\alpha+1}^{\beta-1} \sigma_j \label{R}\\
\sigma_B &= \sum_{j=\beta+1}^{\gamma-1} \sigma_j \label{B}\\
\sigma_L &= \sum_{j=\gamma+1}^{\alpha-1} \sigma_j. \label{L}
\end{align}
\end{subequations}
Notice that Eqs.~(\ref{N_Tri_1})-(\ref{N_Tri_5}) are identical to the win conditions of the triangle game up to some dependence on a ``correction string" given by the outputs of the additional $n-3$ players.  Thus the dependence of the win condition on the correction string does not affect the overall contradictory nature of the system of equations.  Indeed, summing Eqs.~(\ref{N_Tri_1})-(\ref{N_Tri_5}) returns $\sum_{j\in\{\alpha,\beta,\gamma\}} (d_j + b_j + e_j) + \sum_{\Sigma\in\{R,B,L\}} (a_\Sigma + b_{\Sigma} + c_\Sigma) =1$, implying that the total parity of all player's outputs for input $(1,1,1)$ cannot have even parity.  Thus, no classical strategy where players 
do not communicate can win the multiplayer triangle game with probability greater than 7/8.

For this multiplayer game, even classical strategies assisted by geometrically local communication between players fail to win with probability greater than $7/8$.  Consider communication-assisted classical strategies in which players $\alpha$, $\beta$, and $\gamma$ communicate with mutually exclusive parties.  With this restriction the outputs of the players are restricted to be affine boolean functions of the input.  Furthermore, suppose that players $\alpha$, $\beta$, and $\gamma$ communicate only with players on adjacent edges (i.e. player $\alpha$ cannot communicate with players in $B$, player $\beta$ cannot communicate with players in $L$, and player $\gamma$ cannot communicate with players in $R$).  In this case the affine boolean function for the collective outputs of players on each edge of the triangle are independent of the opposite input (i.e. $\sigma_R = \sigma_R(x_\alpha,x_\beta)$, $\sigma_B = \sigma_B(x_\beta,x_\gamma)$, and $\sigma_L = \sigma_L(x_\alpha,x_\gamma)$).  Substituting these Boolean functions into Eqs.~(\ref{N_Tri_1})-(\ref{N_Tri_5}) reveals that the system of equations are still contradictory.  

If the communication can be geometrically nonlocal, the classical strategy can be perfect.  For example, the strategy attributed to the boolean functions; $c_R(\mathbf{x}) = x_\alpha + x_\gamma$, $c_B(\mathbf{x}) = x_\alpha + x_\beta$, $c_L(\mathbf{x}) = x_\beta + x_\gamma$, and all other outputs being 0, satisfies the win conditions.  To see how quantum strategies can surpass even nonlocal communication-assisted classical strategies, it is imperative to move to a 2D scenario on a lattice in which there are many possible cycles between the three players.  For communication-assisted classical strategies with a constant number of rounds of communication between at most $K$ players at a time, it becomes increasingly likely that there will be a cycle in the lattice satisfying the locality conditions that ensure failure of the classical strategy.  We now elaborate on this 2D setting, which gives a worst-case unconditional separation between constant-depth classical and quantum circuits \cite{bravyi2018quantum}.

\subsection{Physical setting for demonstrating unconditional separation}
For completeness we first restate the informal version of Cor.~\ref{SPTO_Advantage}

\begin{SPTO_Advantage}
$(\mathrm{Informal})$ Consider a relation problem given by the multiplayer triangle game embedded on arbitrary cycles in a 2D lattice.  A quantum device that can prepare a 1D SPTO ground state with string order parameters greater than 1/3 on the arbitrary cycles and perform the fixed-point protocol of Theorem~2 solves the problem with probability greater than 7/8 on all inputs exponentially faster than all classical Boolean circuits.
\label{SPTO_Advantage}
\end{SPTO_Advantage}

Here we expound upon the setting for the computational separation in Cor.~\ref{SPTO_Advantage}.  Consider a black-box device taking inputs and producing outputs in correspondence with vertices in some connected two-dimensional lattice graph $\mathcal{G}=(V,E)$ that can be deformed such that its vertices lie on a grid of size $N\times N$.  The device is then tasked to solve a relation problem called the 2D multiplayer triangle problem, which is an embedding of the multiplayer triangle game into the graph $\mathcal{G}$.  We define the relation problem as follows.

\fbox{\parbox{\textwidth}{
\begin{2D_Triangle}
An input to the problem is provided as a tuple $(\alpha,\beta,\gamma,\mathbf{x},\Gamma)$.  $\alpha$, $\beta$, and $\gamma$ are three arbitrary vertices in the 2D lattice graph $\mathcal{G}$ of size $N \times N$.  $\mathbf{x}\in\{0,1\}^{N^2}$ is any binary string of length $N^2$ and Hamming weight $|\mathbf{x}|\leq 3$, where each bit $x_v$ in the string corresponds to a vertex $v$ in $\mathcal{G}$, and the only possible nonzero bits in $\mathbf{x}$ are $x_\alpha$, $x_\beta$, and $x_\gamma$.  $\Gamma$ is a cycle in $\mathcal{G}$ connecting vertices $\alpha$, $\beta$, and $\gamma$ by paths $\Gamma_{\alpha\beta}$, $\Gamma_{\beta\gamma}$, and $\Gamma_{\gamma\alpha}$.  An output is a $3N^2$-bit string $\mathbf{y}\in\{0,1\}^{3N^2}$ composed of three-bit strings $(y_{0,v},y_{1,v},y_{2,v})$ corresponding to the output of each vertex $v$.   A solution of the problem is a string $\mathbf{y} \in\{0,1\}^{3N^2}$ of even parity satisfying the following linear equations.
\renewcommand{\labelenumi}{$\mathrm{(\Roman{enumi})}$}
\begin{enumerate}
\item \label{One} For all $\mathbf{x}\in\{0,1\}^{N^2}$, $\sum_{v\in \Gamma} y_{1,v} = 0$. 
\item \label{Two} If $(x_\alpha,x_\beta,x_\gamma)=(0,0,0)$, then $\sum_{v\in\Gamma} y_{0,v} = 0$ and $\sum_{v\in\Gamma} y_{2,v} =0$.
\item \label{Three} If $(x_\alpha,x_\beta,x_\gamma)=(1,1,0)$, then $y_{0,\alpha} + y_{2,\beta} + y_{0,\gamma} = 1 + \sum_{v\in \Gamma_{\alpha\beta}} y_{2,v} + \sum_{v\in \Gamma_{\beta\gamma}\cup \Gamma_{\gamma\alpha}} y_{0,v}$. 
\item \label{Four} If $(x_\alpha,x_\beta,x_\gamma)=(0,1,1)$, then $y_{0,\alpha} + y_{0,\beta} + y_{2,\gamma} = 1 + \sum_{v\in \Gamma_{\beta\gamma}} y_{2,v} + \sum_{v\in \Gamma_{\alpha\beta}\cup \Gamma_{\gamma\alpha}} y_{0,v}$. 
\item \label{Five} If $(x_\alpha,x_\beta,x_\gamma)=(1,0,1)$, then $y_{2,\alpha} + y_{0,\beta} + y_{0,\gamma} = 1 + \sum_{v\in \Gamma_{\gamma\alpha}} y_{2,v} + \sum_{v\in \Gamma_{\alpha\beta}\cup \Gamma_{\beta\gamma}} y_{0,v}$.
\end{enumerate}
\end{2D_Triangle}
}}

Note that each input specifies an instance of the multiplayer triangle game on the cycle $\Gamma\subset\mathcal{G}$.  The vertices $\alpha$, $\beta$, and $\gamma$ correspond to the players at the corners of the triangle and each vertices' output $(y_{0,v},y_{1,v},y_{2,v})$ around the cycle corresponds to the three-bit output string $\mathbf{y}_j$ of player $j$.  Furthermore, the defining linear relations are equivalent to Eqs.~(\ref{N_Tri_1})-(\ref{N_Tri_5}).  

A key insight of \cite{bravyi2018quantum} is that communication-assisted classical strategies for multiplayer non-local games are in one to one correspondence with classical Boolean circuits. The analogy goes as follows.
\begin{itemize}
\item[(1)] Each player is represented by a spatial block of polynomially many ``wires" in the circuit.  The inputs to these wires can be initialized to be the player's given input $x_j$, an ancillary input $0$, or a random bit drawn from some distribution.
\item[(2)] Communication between $K$ players is represented in the circuit by a gate with fan-in $K$ and unbounded fan-out. When $K$ players $j_1,\cdot\cdot\cdot,j_K$ communicate they share their inputs $x_{j_1},\cdot\cdot\cdot,x_{j_K}$ with each other.  With this knowledge, each player may compute arbitrary Boolean functions $f(x_{j_1},\cdot\cdot\cdot,x_{j_K})$, of those inputs.  The gate computes this Boolean function.  The inputs may be reused as many times as desired due to the unbounded fan-out.
\item[(3)] Each player's output $\mathbf{y}_j$ is represented by the output of three chosen wires in their respective spatial block of wires.
\end{itemize}

The device is a \textit{constant-depth classical device} or \textit{$\mathsf{NC}^0$-device} if internally it can perform classical circuits of constant-depth consisting of arbitrarily nonlocal gates with bounded fan-in at most $K$.  Likewise, the device is a \textit{constant-depth quantum device} or $\mathsf{QNC}^0$-\textit{device} if internally it can perform constant-depth quantum circuits consisting of quantum gates with bounded fan-in at most $K$.  As summarized in the formal version of Cor.~1 below, a constant-depth quantum device can solve the relation problem with probability greater than $7/8$ for all possible inputs.  Indeed, it is possible to prepare fixed-point SPTO ground states in constant depth when the symmetry $G$ is disregarded \cite{chen2010local}. Generic states in the SPTO phase can be prepared well approximately in the constant-depth circuit that simulates symmetric, quasi-adiabatic continuation of Hamiltonians from the fixed-point \cite{hastings2005quasiadiabatic}. Thus, the advantageous quantum strategy is achieved by first constructing a 1D SPTO ground state with sufficiently large string order parameter $\langle\mathcal{S}\rangle>1/3$ and appropriate symmetry group---for example, $\mathbb{Z}_2\times\mathbb{Z}_2$---and then implementing the respective fixed-point protocol as in Thm.~2.  In other words, each player holds $l$ constituent particles of the 1D SPTO chain and measures in one of the horizontal or vertical context of the table
\begin{align}
\vcenter{\hbox{\includegraphics[width=0.275\linewidth]{Triangle_Square}}}
\label{Tri_Sqr}
\end{align}
depending on if their received input is 0 or 1, respectively.

This quantum strategy can be implemented by constant-depth quantum circuits that are also geometrically local with respect to the graph $\mathcal{G}$. However, we will prove in the following that a constant-depth classical device cannot solve this relation problem with probability greater than $7/8$ on all inputs.  

Notice that the set of all inputs $\mathcal{I}= \{(\alpha,\beta,\gamma,\mathbf{x},\Gamma)\}$ has size $|\mathcal{I}|=O(2^{N^2})$, since there are an exponential number of cycles in any 2D lattice graph.  However, to demonstrate the quantum computational advantage, we only need to check a polynomially large subset of provably hard instances $\mathcal{I}_{\textrm{Hard}}\subset \mathcal{I}$ of size $|\mathcal{I}_{\textrm{Hard}}| = O(N^{16/3})$.  An exact description of $\mathcal{I}_{\textrm{Hard}}$ and the proof of its polynomial size are given in the next section. It will be shown that any classical circuit that solves the problem with probability greater than $7/8$ on this subset of inputs must have depth $D\geq \frac{2}{5}\log_K(N)$ (i.e. depth that is logarithmic in the system size).  This demonstrates an unconditional exponential separation between the power of quantum and classical circuits. In particular, it implies a worst-case separation of computational complexity classes $\mathsf{NC}^0\subsetneq\mathsf{QNC}^0$.

\begin{figure}
\includegraphics[width=\linewidth]{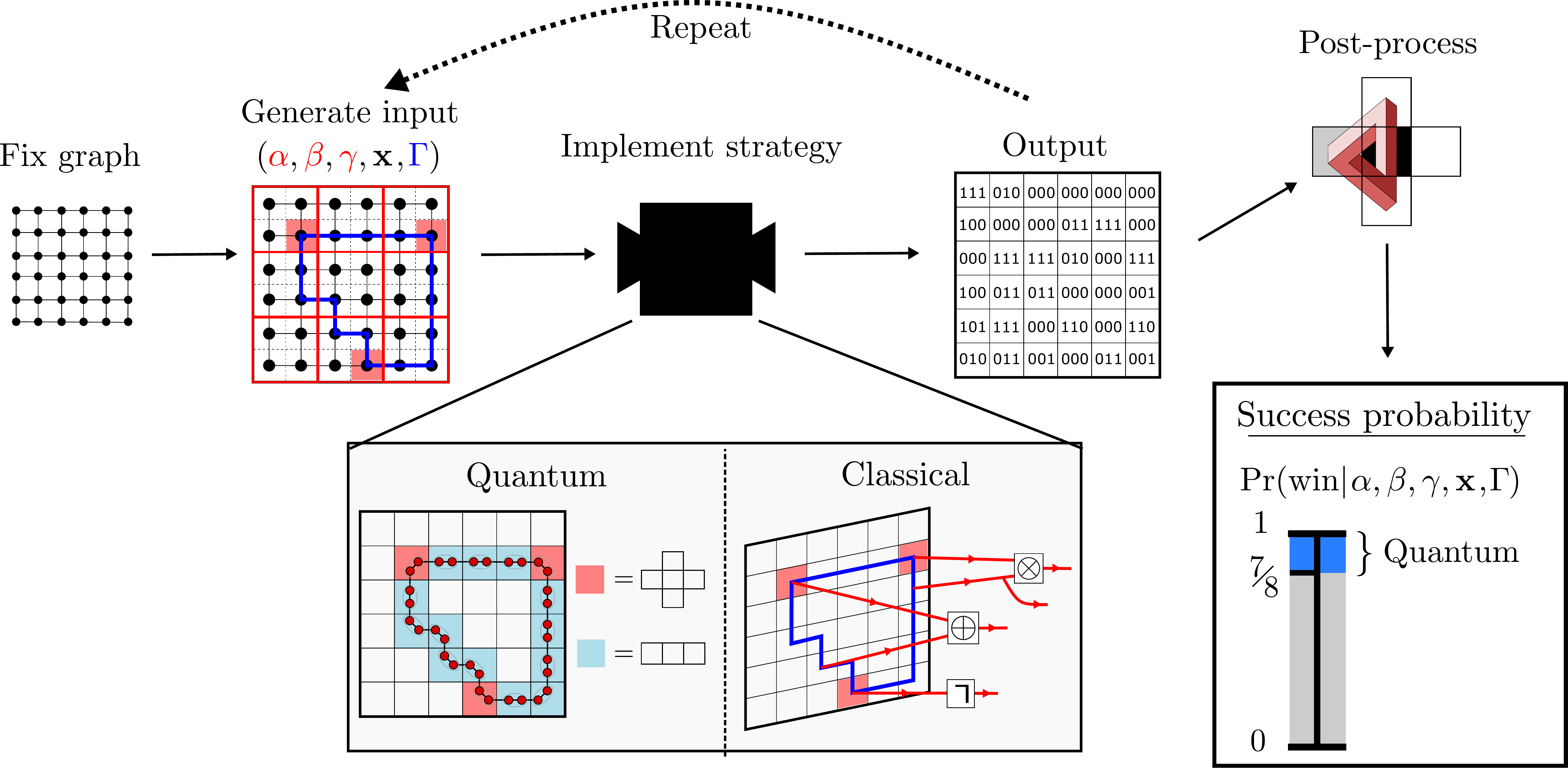}
\caption{Setting for the unconditional exponential separation between classical and quantum circuits.  For a fixed lattice graph $\mathcal{G}$, depicted here as the grid graph, a blackbox device is tasked to solve the 2D multiplayer triangle problem on inputs drawn from $\mathcal{I}_\textrm{Hard}$.  Each input is a tuple $(\alpha,\beta,\gamma,\mathbf{x},\Gamma)$.  $\alpha$, $\beta$, and $\gamma$ are three vertices depicted by the shaded red squares chosen from three distinct boxes, outlined in red.  $\mathbf{x}\in\{0,1\}^{N^2}$ is a binary string of Hamming weight $|\mathbf{x}|\leq 3$ whose nonzero inputs correspond to the vertices $\alpha$, $\beta$, and $\gamma$.  $\Gamma$ is a cycle connecting $\alpha$, $\beta$, and $\gamma$, depicted in blue.  A randomly generated input is given to a constant-depth quantum device that implements the 1D SPTO strategy for the multiplayer triangle game on the cycle $\Gamma$.  The performance of the quantum device is compared against all classical strategies, which can implement classical circuits consisting of AND, XOR, and NOT gates (denoted $\otimes$, $\oplus$, and $\neg$) with arbitrary locality and fan-in at most $K$.  The device produces an output, which consists of a three-bit string for each vertex in $\mathcal{G}$.  The outputs are then checked against the linear relations defining the solution set.  Repeating this many times for many different inputs, the success probability for each input is obtained.  If for each input in $\mathcal{I}_{\textrm{Hard}}$ the success probability is greater than $7/8$, the constant-depth quantum device outperforms exponentially all classical circuits as the latter needs at least logarithmic depth to match.}
\label{Advantage_Setting}
\end{figure}

The protocol by which the unconditional separation can be demonstrated is depicted in Fig.~\ref{Advantage_Setting}.  For a fixed lattice graph $\mathcal{G}$, generate a random input drawn uniformly from the subset of hard instances, $(\alpha,\beta,\gamma,\mathbf{x},\Gamma)\in\mathcal{I}_{\textrm{Hard}}$.  Given the input, the device implements the quantum strategy for the multiplayer triangle game on the cycle $\Gamma$ and outputs a three-bit string for each vertex in $\mathcal{G}$.  The outputs are then checked against the linear relations (\ref{One})-(\ref{Five}) and the result is recorded as a success or failure. This procedure is repeated many times to build up statistics for the success probability for each input.  If for all inputs the success probability is greater than $7/8$, then the computational separation has been demonstrated.

\subsection{Proof of unconditional separation}
\label{Proof_of_separation}
In this section we give a formal proof of Cor.~\ref{SPTO_Advantage} from the main text.  Given Theorem~2, the formal proof does not need any radically new ideas beyond the results of \cite{bravyi2018quantum, coudron2018trading, gall2018average}.

As described above, communication-assisted classical strategies for the multiplayer triangle game that simply compute affine Boolean functions with geometrically restricted dependence succeed with probability no greater than $7/8$ (i.e. they must fail for one or more inputs).  We first recast this condition for failure in terms of correlations generated by a classical circuit, which are conveniently expressed via \textit{light cones}.
\begin{Definition}
\cite{bravyi2018quantum, gall2018average} Let $\mathcal{C}$ be a classical circuit with inputs indexed by set $\mathbf{X}$ and outputs indexed by set $\mathbf{Y}$.  Given an input $x\in \mathbf{X}$, the forward light cone of $x$, denoted $L^+_\mathcal{C}(x)$, is the set of all output bits that depend on $x$.  Similarly, the backward light cone of an output $y\in \mathbf{Y}$, denoted $L^-_{\mathcal{C}}(y)$, is the set of all inputs it depends on.
\end{Definition}
With this definition we can recast the conditions for failure in terms of conditions on the light cones of the inputs.
\fbox{\parbox{\textwidth}{
\begin{Failure}
For a given input $(\alpha,\beta,\gamma,\mathbf{x},\Gamma)$ for the 2D multiplayer triangle problem, a classical circuit $\mathcal{C}$ will fail to solve the problem for at least one string $\mathbf{x}$ whenever the following conditions hold.
\begin{itemize}
\item[$\mathrm{(I)}$] The forward light cones of the possible nonzero inputs $x_\alpha$, $x_\beta$, and $x_\gamma$ are pairwise disjoint.
\begin{align}
L_{\mathcal{C}}^+(x_\alpha)\cap L_{\mathcal{C}}^+(x_\beta) = L_{\mathcal{C}}^+(x_\alpha)\cap L_{\mathcal{C}}^+(x_\gamma) = L_{\mathcal{C}}^+(x_\beta)\cap L_{\mathcal{C}}^+(x_\gamma)   = \varnothing. \label{Fail1}
\end{align}
\item[$\mathrm{(II)}$] Let $\Gamma_{ab}$ denote the direct path within the cycle going from vertex $a$ to vertex $b$.  The forward light cones of each possibly nonzero input are disjoint from the outputs on the opposite edge.  I.e.
\begin{subequations}
\begin{align}
L_{\mathcal{C}}^+(x_\alpha)\cap\Gamma_{\beta\gamma} &= \varnothing \label{Fail2}\\
L_{\mathcal{C}}^+(x_\beta)\cap\Gamma_{\alpha\gamma} &= \varnothing \label{Fail3}\\
L_{\mathcal{C}}^+(x_\gamma)\cap\Gamma_{\alpha\beta} &= \varnothing \label{Fail4}.
\end{align}
\end{subequations}
\end{itemize}
\end{Failure}
}}
We now show that for sufficiently large $N$ any classical circuit of depth $D<\frac{2}{5}\log_K(N)$ will satisfy the above failure conditions for at least one input $(\alpha,\beta,\gamma,\mathbf{x},\Gamma)\in\mathcal{I}_{\textrm{Hard}}$.

\begin{Proposition}
\cite{bravyi2018quantum, gall2018average} For any classical circuit $\mathcal{C}$ of depth $D$ consisting of gates with fan-in at most $K$, all outputs $y\in \mathbf{Y}$ have backwards light cones of bounded size $|L_{\mathcal{C}}^-(y)|\leq K^D$.
\label{Prop1}
\end{Proposition}
\textit{Proof}: For each layer of gates each output is correlated with at most $K$ inputs.  Therefore, after $D$ layers of gates each output is correlated with at most $K^D$ inputs.  $\square$

If the depth of the circuit satisfies $D<\frac{2}{5}\log_K(N)$, we find $|L_{\mathcal{C}}^-(y)|\leq N^{2/5}$.  Let $V_{\textrm{Big}}=\{x\in\mathbf{X}~|~|L^+(x)|\geq N^{1/2}\}$ and $V_{\textrm{Small}}=\{x\in\mathbf{X}~|~|L^+(x)|<N^{1/2}\}$.

\begin{Proposition}
\cite{bravyi2018quantum, gall2018average} $|V_{\mathrm{Big}}|<N^{19/10}$.
\label{Prop2}
\end{Proposition}
\textit{Proof}: Consider the interaction graph $\mathcal{G}_{\textrm{int},\mathcal{C}}$ of the circuit $\mathcal{C}$ defined as the bipartite graph with vertex set $V_{\textrm{int}}=\mathbf{X}\cup\mathbf{Y}$ and edge set $E_{\textrm{int}}$ where $(x,y)\in E_{\textrm{int}}$ iff $x\in L_{\mathcal{C}}^-(y)$. $|E_{\textrm{int}}|$ satisfies, $|V_{\textrm{Big}}| N^{1/2}<|E_{\textrm{int}}|<N^2 N^{2/5}$ which implies that $|V_{\textrm{Big}}|<N^{19/10}$ so $|V_{\textrm{Small}}|=\Omega(N^2)$ (i.e. most forward light cones are small). $\square$

Now consider the lattice graph $\mathcal{G}=(V,E)$ over which the classical circuit $\mathcal{C}$ takes inputs and generates outputs.  This graph can be partitioned into $N^{2/3}$ disjoint connected sets containing $N^{4/3}$ vertices with diameter $\Theta(N^{2/3})$, where the diameter is defined with respect to the Euclidean distance on the grid, referred to as neighborhoods. For each vertex $v\in V$ let $\textrm{Nbhd}(v)$ denote the unique neighborhood to which that vertex belongs.  Also define three connected regions $\mathcal{U},\mathcal{V},\mathcal{W}\subset V_{\textrm{Small}}$, each containing $\left\lfloor \frac{N}{3}\right\rfloor^2$ vertices, that are well separated from each other in that $\textrm{dist}(\mathcal{U},\mathcal{V}),\textrm{dist}(\mathcal{U},\mathcal{W}),\textrm{dist}(\mathcal{V},\mathcal{W})\geq \left\lfloor\frac{N}{3}\right\rfloor$.  Furthermore, the diameter of these sets is $\textrm{diam}(\mathcal{U}),\textrm{diam}(\mathcal{V}),\textrm{diam}(\mathcal{W})=\Theta(N)$.

The following proposition gives a bound on the probability that a randomly selected triple of vertices $(\alpha,\beta,\gamma)$ with $\alpha\in\mathcal{U}$, $\beta\in\mathcal{V}$, and $\gamma\in\mathcal{W}$ have forward light cones that intersect each other's neighborhoods.
\begin{Proposition}
\cite{bravyi2018quantum, gall2018average} Let $(\alpha,\beta,\gamma)$ be a randomly selected triple of vertices with $\alpha\in\mathcal{U}$, $\beta\in\mathcal{V}$, and $\gamma\in\mathcal{W}$.  Consider the following set of light cone conditions,
\begin{subequations}
\begin{align}
&L_{\mathcal{C}}^+(x_\alpha)\cap\mathrm{Nbdh}(\beta) = \varnothing \textrm{ and } L_{\mathcal{C}}^+(x_\alpha)\cap\mathrm{Nbdh}(\gamma) = \varnothing \label{Nbdh_1}\\
&L_{\mathcal{C}}^+(x_\beta)\cap\mathrm{Nbdh}(\alpha) = \varnothing \textrm{ and } L_{\mathcal{C}}^+(x_\beta)\cap\mathrm{Nbdh}(\gamma) = \varnothing \label{Nbdh_2}\\
&L_{\mathcal{C}}^+(x_\gamma)\cap\mathrm{Nbdh}(\alpha) = \varnothing \textrm{ and } L_{\mathcal{C}}^+(x_\gamma)\cap\mathrm{Nbdh}(\beta) = \varnothing.\label{Nbdh_3}
\end{align}
\end{subequations}
The probability that any one of the following conditions fails to hold is $O(N^{-1/6})$.  Therefore, for sufficiently large $N$ there will be at least one triple for which Eqs.~(\ref{Nbdh_1})-(\ref{Nbdh_3}) are satisfied.
\label{Prop3}
\end{Proposition}
\textit{Proof}: Without loss of generality consider vertex $\alpha\in\mathcal{U}$ and its corresponding input $x_\alpha$.  Since $\alpha\in V_{\textrm{Small}}$ its forward light cone $L_\mathcal{C}^+(x_\alpha)$ intersects at most $N^{1/2}$ different neighborhoods.  Since each neighborhood contains $N^{4/3}$ vertices, the total number of vertices $v\in V$ satisfying $L_\mathcal{C}^+(x_\alpha)\cap\textrm{Nbdh}(v)\neq\varnothing$ is $O(N^{11/6})$.  Since the total number of vertices in $\mathcal{V}$ is $O(N^2)$ we have that $\textrm{pr}[L_{\mathcal{C}}^+(x_\alpha)\cap\mathrm{Nbdh}(\beta) \neq \varnothing]= O(N^{-1/6})$.  By the union bound the probability that any one of Eqs.~(\ref{Nbdh_1})-(\ref{Nbdh_3}) is not satisfied is $O(N^{-1/6})$. $\square$

The following proposition deals with the likelihood that the failure condition (I), Eq.~(\ref{Fail1}), is satisfied for some input.
\begin{Proposition}
\cite{bravyi2018quantum, gall2018average} Let $(\alpha,\beta,\gamma)$ be a triple of vertices as described above.  The probability that any one of the equalities in Eq.~(\ref{Fail1}) fails to hold is $O(N^{-3})$.  Therefore, for sufficiently large $N$ there will be at least one triple for which Eq.~(\ref{Fail1}) is satisfied.
\label{Prop4}
\end{Proposition}
\textit{Proof}: Without loss of generality consider vertices $\alpha$ and $\beta$.  The probability that a vertex $v\in V$ lies in $L_\mathcal{C}^+(x_\alpha)$ is $O(N^{-3/2})$.  Thus, $\textrm{pr}[v\in L_\mathcal{C}^+(x_\alpha) \textrm{ and } v\in L_\mathcal{C}^+(x_\beta) ] = O(N^{-3})$. By the union bound the probability that the forward light cones of any pair of vertices from the triple $(\alpha,\beta,\gamma)$ intersect is $O(N^{-3})$.  $\square$

The following proposition deals with the likelihood that the failure condition (II), Eqs.~(\ref{Fail2})-(\ref{Fail4}), are satisfied for some input.
\begin{Proposition}
\cite{bravyi2018quantum, gall2018average} For the triple $(\alpha,\beta,\gamma)$ described above, construct a random cycle $\Gamma$ formed by three direct paths between each pair of vertices $\Gamma_{\alpha\beta}$, $\Gamma_{\beta\gamma}$, and $\Gamma_{\alpha\gamma}$.  The probability that Eqs.~(\ref{Fail2})-(\ref{Fail4}) are not satisfied is $O(N^{-1/6})$.   Therefore, for sufficiently large $N$ there will be at least one cycle $\Gamma$ for which Eqs.~(\ref{Fail2})-(\ref{Fail4}) are satisfied.
\label{Prop5}
\end{Proposition}
\textit{Proof}: Without loss of generality consider the path $\Gamma_{\alpha\beta}$.  Since the triple $(\alpha,\beta,\gamma)$ satisfies Eqs.~(\ref{Nbdh_1})-(\ref{Nbdh_3}) we need only consider the section of the path $\Gamma_{\alpha\beta}'=\Gamma_{\alpha\beta}\backslash (\textrm{Nbdh}(\alpha)\cup\textrm{Nbdh}(\beta))$, which extends from the boundary of $\textrm{Nbdh}(\alpha)$ to the boundary of $\textrm{Nbdh}(\beta)$.  We can construct $O(N^{2/3})$ disjoint paths $\Gamma_{\alpha\beta}'$, which emerge from half of the points on the circumference of $\textrm{Nbdh}(\alpha)$ and $\textrm{Nbdh}(\beta)$. Since $|L_\mathcal{C}^+(x_\gamma)|<N^{1/2}$ it can at most intersect $N^{1/2}$ such paths $\Gamma_{\alpha\beta}'$.  Thus for a randomly chosen path $\Gamma_{\alpha\beta}$, $\textrm{pr}[L_\mathcal{C}^+(x_\gamma)\cap\Gamma_{\alpha\beta}\neq\varnothing]=O(N^{-1/6})$.  By the union bound the probability that any one of Eqs.~(\ref{Fail2})-(\ref{Fail4}) are not satisfied is $O(N^{-1/6})$.  $\square$

Propositions \ref{Prop1}--\ref{Prop5} define a subset $\mathcal{I}_{\textrm{Hard}}\subset\mathcal{I}$ of the input set that are provably hard for constant depth classical circuits.  We now show that  $\mathcal{I}_{\textrm{Hard}}$ has size growing polynomially in the grid size.
\begin{Proposition}
$|\mathcal{I}_{\textrm{Hard}}|=O(N^{16/3})$.
\label{Prop6}
\end{Proposition}
\textit{Proof:}. Divide $\mathcal{G}$ into $N^{2/3}$ disjoint contiguous regions (neighborhoods) each containing $N^{4/3}$ vertices and each having diameter $O(N^{2/3})$. For example, one could use ``boxes" of size $N^{2/3}\times N^{2/3}$ in the $N\times N$ square grid as in Ref.~\cite{bravyi2018quantum}.  To obtain $\alpha$, $\beta$, and $\gamma$ choose three vertices from three distinct neighborhoods (there are $O(N^{10/3})$ such choices).  $\mathbf{x}$ is obtained by choosing from one of the eight possible strings with appropriate support.  Finally, for each pair of neighborhoods choose one path connecting them from a set of $O(N^{2/3})$ non-intersecting paths between the boundary of each.  For each choice of three paths we can connect their endpoints to the nearest vertex $\alpha$, $\beta$, or $\gamma$ via a path within each neighborhood to create a cycle $\Gamma$.  There are $O(N^{2})$ such cycles.  Therefore, subset of hard instances has polynomial size $|\mathcal{I}_{\textrm{Hard}}|=O(N^{16/3})$. $\square$

\begin{Cor}
$\mathrm{(Formal)}$ Consider the 2D multiplayer triangle problem on a 2D lattice graph $\mathcal{G}$ of size $N \times N$.  Consider a constant-depth quantum device with two capabilities (i) to prepare a generic 1D SPTO ground state on arbitrary cycles in $\mathcal{G}$ with a string order parameter greater than $\frac{1}{3}$, and (ii) to measure sites in the contexts of Eq.~(\ref{Tri_Sqr}), performing the fixed-point protocol of Thm.~2.  This quantum device outputs a solution of the 2D multiplayer triangle problem with probability higher than $\frac{7}{8}$ for all the inputs, which any classical circuit of depth $D<\frac{2}{5}\log_K(N)$ cannot do.
\end{Cor}
\textit{Proof}: By Props.~\ref{Prop1}-\ref{Prop5}, for any classical circuit consisting of gates with fan-in $K$ and depth $D<\frac{2}{5}\log_{K}(N)$ there exists at least one input $(\alpha,\beta,\gamma,\mathbf{x},\Gamma)$ satisfying the failure conditions of Eq.~(\ref{Fail1}) and Eqs.~(\ref{Fail2})-(\ref{Fail4}).  Therefore, such a classical circuit cannot succeed on all inputs with probability greater than $7/8$.  On the other hand, a constant-depth quantum device with the above capability can.  Therefore, the constant-depth quantum device outperforms all such sub-logarithmic depth classical circuits.  $\square$

\section{Determination of the boundary operators $V^L(g)$ and $V^R(g)$}
\label{SM2}

In this section we will review matrix product state representations of 1D SPTO ground states and use them to construct the fixed-point boundary operators, $V^L(g)$ and $V^R(g)$, used in the text.  We will use these explicit expressions to prove various properties of the boundary operators from the main text.  Namely, we show the following properties of the fixed-point boundary operators
\begin{align}
V^R(g)V^R(h) &= \omega(g,h) V^R(gh) \label{R_Proj} ,\\ 
V^L(g)V^L(h) &= \omega(g,h)^* V^L(gh) \label{L_Proj} ,\\
V^R(g)V^L(h) &= V^L(h)V^R(g) \label{RL_Com} ,\\
V^R(g)V^L(g)&= u(g)^{\otimes l} \label{RL_Sym}.
\end{align}
Furthermore, for a fixed-point ground state $|\psi\rangle$
\begin{align}
S_{[j,k]}(g)|\psi\rangle &= \left(V^L_j(g)\otimes V^R_k(g)\right)U_{[j,k]}(g)|\psi\rangle = |\psi\rangle \label{Trunc_Sym} \\
T_{[j,k]}^{(g,h)} |\psi\rangle &= V_{k}^R(g) U(h) V_{j}^L(g) U_{[j,k]}(g)|\psi\rangle = \Omega(g,h) |\psi\rangle. \label{Twist_Phase_Expectation}
\end{align}

\subsection{Matrix product states and tensor network notation}

A matrix product state (MPS) representation of a quantum state expresses the amplitude of each basis vector as the trace of a product of matrices.  In particular we write,
\begin{align}
|\psi\rangle = \sum_{j_0,...,j_{N-1}} \textrm{Tr}\left(A^{(j_{N-1})}\cdot\cdot\cdot A^{(j_0)}\right) |j_0, ..., j_{N-1}\rangle
\end{align}
for uniform many-body states on periodic boundary conditions, where $A^{(j_k)}\in L(\mathbb{C}^D)$ is a linear operator on the vector space $\mathbb{C}^D$ (i.e. a complex $D\times D$ matrix).  The vector space, $\mathbb{C}^D$, on which the matrices act is referred to as the virtual space.  Its dimension is called the bond dimension of the MPS.  

An MPS can seen as a contraction of many 3-index tensors,
\begin{align}
\mathcal{A} = \sum_{j=0}^{d-1}\sum_{k,k'=0}^{D-1} A^{(j)}_{k,k'} |j;k\rangle\langle k'|.
\end{align}
When studying MPS we will appeal to the diagrammatic tensor network notation for preforming calculations.  In tensor network notation multi-index objects are denoted as boxes with lines emerging from them.  Each line represents an index that takes values in the corresponding indexing set.  Connecting two lines denotes a contraction over those indices. In tensor network notation the MPS is expressed as,
\begin{align}
|\psi\rangle &= \vcenter{\hbox{\includegraphics[width=0.3\linewidth]{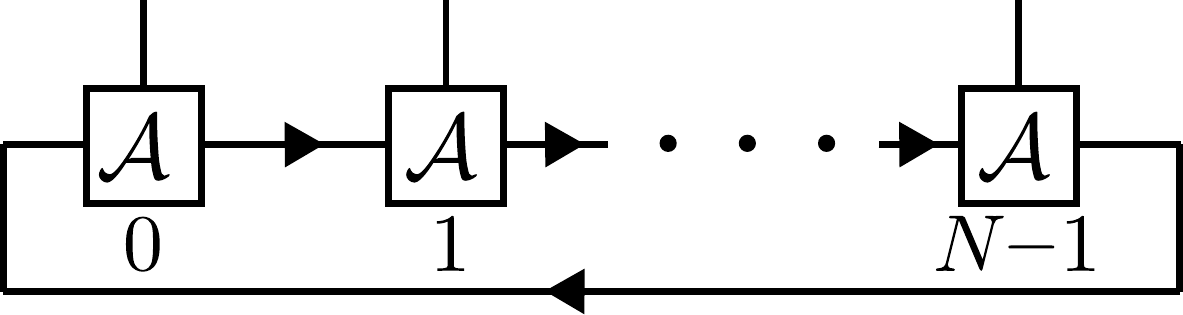}}}.
\end{align}
In the diagrammatic representation we have labeled the lines representing the virtual space with arrows directed in a clockwise manner.  These arrows are used to depict the order in which multiplication should be performed.  Here we will work with the convention that, multiplication is to be performed in the opposite order as denoted by the arrow.  For instance,
\begin{align}
\vcenter{\hbox{\includegraphics[width=0.13\linewidth]{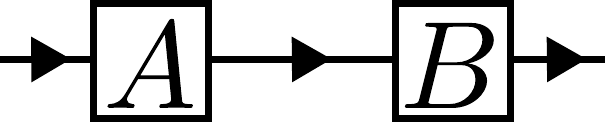}}} &= \vcenter{\hbox{\includegraphics[width=0.08\linewidth]{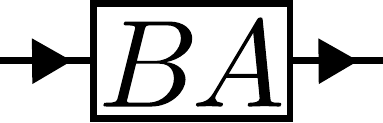}}}.
\end{align}
For a good review of tensor network notation see \cite{bridgeman2017hand}.

\subsection{Matrix product states and symmetry-protected topological order}

1D SPTO phases can be classified using MPS according their symmetry properties \cite{schuch2011classifying}.   Consider symmetry groups that carry a uniform \textit{on-site} unitary representation $U(g) = u(g)^{\otimes N}$, i.e. represented in tensor product on the many-body Hilbert space $\mathcal{H} = (\mathbb{C}^d)^{\otimes N}$.  The on-site representation $u(g)$ maps each component of the MPS tensor $A^{(j)}$ to $\sum_{k}u(g)_{jk}A^{(k)}$.  Since the state is invariant under the global symmetry, the transformed MPS tensors must be related to the original ones by a ``gauge transformation" \cite{perez2007matrix}
\begin{align}
\sum_{k=0}^{d-1}u(g)_{jk}A^{(k)} = V(g)^\dagger A^{(j)}V(g).
\end{align}
In tensor network notation we say that the on-site representation $u(g)$ can be \textit{pushed through} each local MPS tensor $\mathcal{A}$ to yield another representation $V(g)$, which acts via conjugation on the virtual space.  In tensor network notation this is written,
\begin{align}
\vcenter{\hbox{\includegraphics[width=0.3\linewidth]{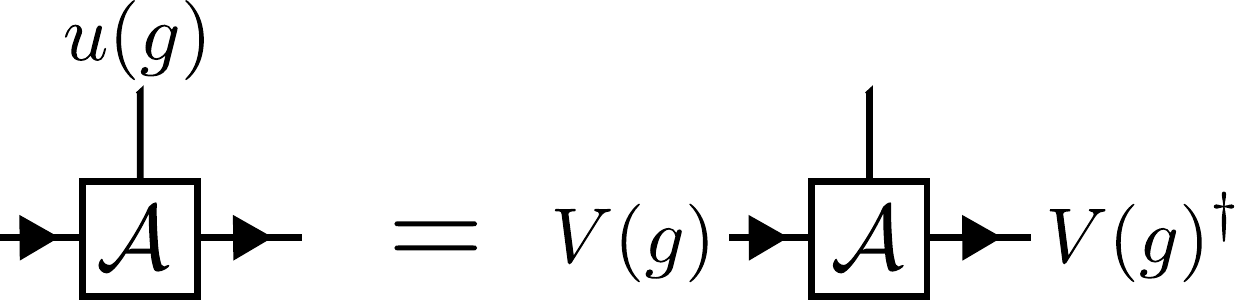}}}.  \label{MPS_Prep}
\end{align}
Symmetric MPS represent 1D SPTO ground states whenever the unitaries $V(g)$ form a nontrivial \textit{projective representation} as described in the text.  

Algebraically, a projective representation of $G$ is a map $V:G\rightarrow GL(\mathbb{V})$, for some vector space $\mathbb{V}$, 
that obeys the group multiplication law up to a $G$-dependent phase, i.e.,
\begin{align}
V(g)V(h) = \omega(g,h)V(gh),
\end{align}
where $\omega(g,h)\in U(1)$ is called a 2-cocycle.  
Associativity is assured by the 2-cocycle condition, $\omega(a,b)\omega(ab,c) = \omega(a,bc)\omega(b,c)$ $\forall a,b,c\in G$.  A projective representation is said to be nontrivial whenever the 2-cocycle cannot be removed by a multiplicative phase, $d\beta(g,h) = \beta(g)\beta(h)/\beta(gh)$, known as a 2-coboundary.  This defines an equivalence relation on 2-cocycles where $\omega\sim\omega'$ iff $\omega'=(d\beta)\omega$.  
The equivalence classes, denoted $[\omega]$, form a multiplicative group called the second group cohomology, $H^2(G,U(1))$.  1D SPTO phases are in one to one correspondence with elements of $H^2(G,U(1))$ \cite{chen2011classification}.

\subsection{The boundary operators in the MPS representation}

The operators $V^L(g)$ and $V^R(g)$ introduced in the text send the state resulting from a truncated symmetry transformation back to the ground state as described in Eq.~(\ref{Trunc_Sym}).  The same action can be achieved in the virtual space by placing $V(g)^\dagger$ and $V(g)$ on the virtual indices at the left and right endpoints of the truncated symmetry operator.  We thus expect that $V^L(g)$ and $V^R(g)$ are operators that can be pushed through to the virtual space where they act as $V(g)^\dagger$ and $V(g)$ on the right or left bonds, respectively.

Consider the MPS tensor as a map from the virtual space to the physical space $\mathcal{A}: \mathbb{C}^D\otimes\mathbb{C}^D\rightarrow \mathbb{C}^d$, this map is called the MPS projector.  Given any operator on the virtual space $W$ there is an operator $P$ on the physical space such that $P\mathcal{A} = \mathcal{A}W$ whenever the map $\mathcal{A}$ is injective.  Indeed, the injectivity of $\mathcal{A}$ implies there is a left inverse $\mathcal{A}^{-1}$.  It follows that $P = \mathcal{A} W \mathcal{A}^{-1}$. 

Some MPS projectors $\mathcal{A}$ may not have a left inverse simply because the dimension $d$ of the physical space is simply too small for the map $\mathcal{A}$ to be injective (i.e. $d<D^2$).  Such tensors can be made injective by blocking some number of sites $l$ to form a new tensor $\mathcal{A}^l$ where,
\begin{align}
\mathcal{A}^l = \sum_{j_1,\cdot\cdot\cdot,j_l = 0}^{d-1} \sum_{k,k'=0}^{D-1} \left(\sum_{r_1,\cdot\cdot\cdot,r_{l-1} = 0}^{D-1}A^{(j_{l})}_{k,r_{l-1}}\cdot\cdot\cdot A^{(j_2)}_{r_2,r_1}A^{(j_1)}_{r_1,k'} \right) |j_1,\cdot\cdot\cdot,j_l;k\rangle\langle k'|.
\end{align}
In tensor network notation this is simply written,
\begin{align}
\vcenter{\hbox{\includegraphics[width=0.4\linewidth]{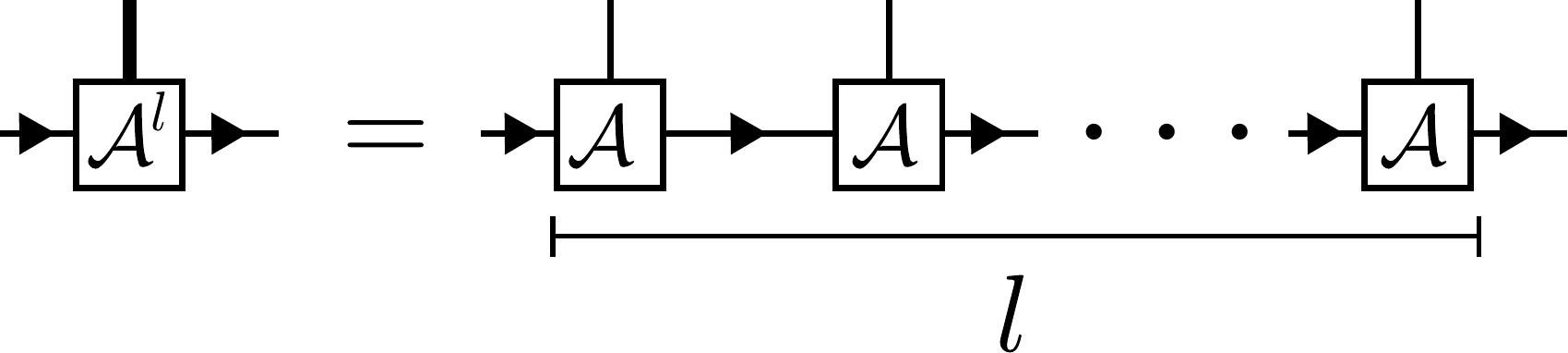}}}.
\end{align}
The thick line on the lefthand side indicates that the physical space dimension is larger.  The left inverse property is written,
\begin{align}
\vcenter{\hbox{\includegraphics[width=0.09\linewidth]{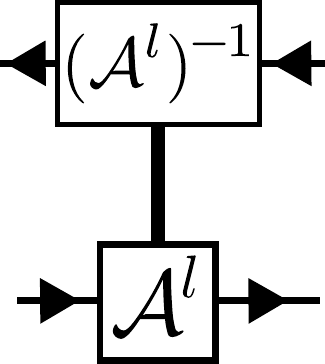}}} &= \vcenter{\hbox{\includegraphics[width=0.06\linewidth]{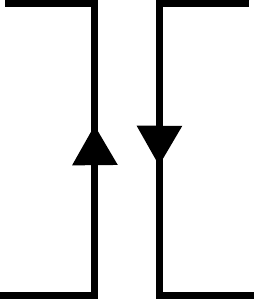}}}. \label{L_Inv_Def}
\end{align}
An MPS whose tensors form injective maps after blocking some number of sites $l$ are called \textit{injective MPS}.  The number of sites to be blocked $l$ is called the \textit{injectivity length}.  It turns out that injectivity is both a necessary and sufficient condition for an MPS to be the unique ground state of its associated parent Hamiltonian.  Thus, MPS representations of 1D SPTO ground states on periodic boundary conditions are always injective MPS.  Furthermore, the projective unitaries can always be found, given the MPS, as,
\begin{align}
V^L(g) &= \mathcal{A}^l(\mathbbm{1}\otimes V(g)^\dagger)({\mathcal{A}^l})^{-1} \\
V^R(g) &= \mathcal{A}^l(V(g)\otimes\mathbbm{1})({\mathcal{A}^l})^{-1}.
\end{align}
In tensor network notation these are,
\begin{align}
V^L(g) &= \vcenter{\hbox{\includegraphics[width=0.1\linewidth]{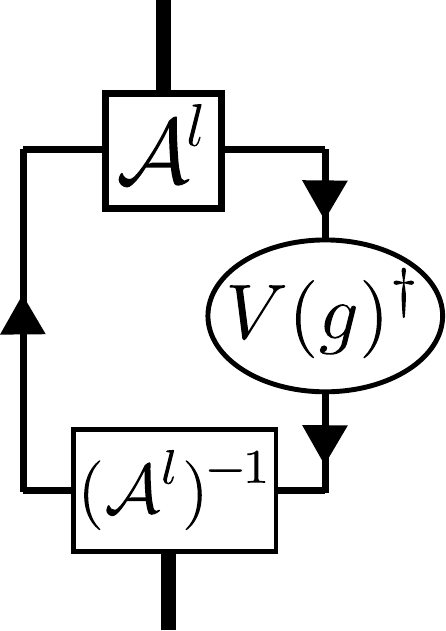}}} \label{VL_TNN} \\
V^R(g) &=  \vcenter{\hbox{\includegraphics[width=0.1\linewidth]{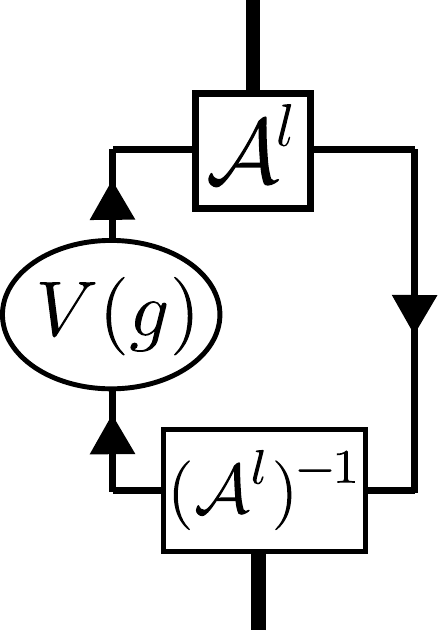}}} \label{VR_TNN} .
\end{align}
These operators have nontrivial support on at most $l$ sites.  In tensor network notation, their action on the MPS can be written,
\begin{align}
&\vcenter{\hbox{\includegraphics[width=0.25\linewidth]{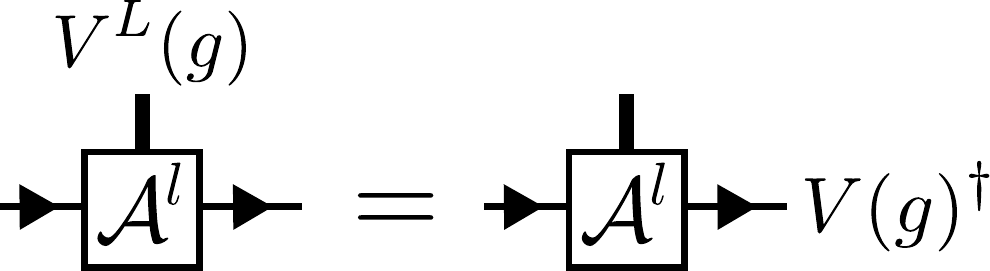}}} \label{V_L_Def} \\
&\vcenter{\hbox{\includegraphics[width=0.24\linewidth]{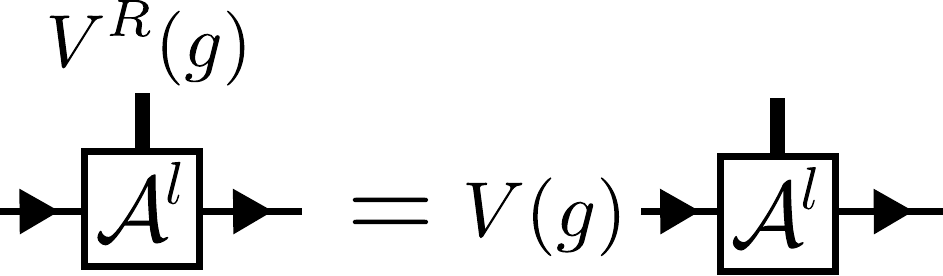} \label{V_R_Def}}}.
\end{align}
Notice that the support of the fixed-point boundary operators used in the main text $m$ is simply the injectivity length of the fixed-point MPS.

The fixed-point boundary operators play a predominant role in the main text.  The fixed-point state is the MPS obtained under quantum state renormalization group (RG) implemented on the \textit{transfer matrix}.  The transfer matrix of a MPS generated by MPS tensor $\mathcal{A} = \sum_{j=0}^{d-1} A^{(j)}\otimes|j\rangle$ is defined as
$\mathcal{E}_{\mathcal{A}}= \sum_{j=0}^{d-1} A^{(j)}\otimes {A^{(j)}}^*.$
This object is also called the ``identity transfer matrix" and will later be denoted as $\mathcal{E}_{\mathbbm{1}}$ for convenience.  The quantum state RG is then implemented by blocking some number $M>1$ transfer matrices and finding a new MPS $\mathcal{A}'$ with equivalent transfer matrix.  In other words, one RG step is implemented by solving for $\mathcal{A}'$ in the equation $\mathcal{E}_{\mathcal{A}'} = \mathcal{E}_{\mathcal{A}}^M$.

The fixed-point obtained under quantum state RG implemented on a given MPS $\mathcal{A}$ can be understood in terms of the left and right eigenvectors of $\mathcal{A}$.  Injective MPS have unique left and right eigenvectors with largest eigenvalue one when written in their so-called canonical form \cite{perez2007matrix}.  The second largest eigenvalue $\lambda_2$  determines the rate of exponential decay of two point correlation functions, which is characterized by the \textit{correlation length} $\xi = 1/\ln(1/\lambda_2)$.  The fixed-point MPS thus has correlation length $\xi=0$. If the left and right eigenvectors are $|R_\mathcal{A}\rangle$ and $|L_\mathcal{A}\rangle$, the fixed-point transfer matrix is $\mathcal{E}_{\mathcal{A}_\textrm{Fix}} = |L_\mathcal{A}\rangle\langle R_{\mathcal{A}}|$.  Thus up to unitary transformation $({\mathcal{A}_{\textrm{Fix}}^l})^{-1}={\mathcal{A}_{\textrm{Fix}}^{l}}^\dagger$ and the fixed point boundary operators are,
\begin{align}
V^L(g) &= \mathcal{A}_{\textrm{Fix}}^l(\mathbbm{1}\otimes V(g)^\dagger){{\mathcal{A}_{\textrm{Fix}}^l}}^\dagger\\
V^R(g) &= \mathcal{A}_{\textrm{Fix}}^l(V(g)\otimes\mathbbm{1}){{\mathcal{A}_{\textrm{Fix}}^l}}^\dagger.
\end{align}
The fixed-point boundary operators are thus guaranteed to be unitary and can thus be measured via a projective value measure (PVM).

\subsection{Properties of the boundary operators in the MPS representation}
We now prove each expression.
 \begin{itemize}
 \item We first prove Eq.~(\ref{Trunc_Sym}).  This follows simply from Eqs.~(\ref{MPS_Prep}), (\ref{V_L_Def}), and (\ref{V_R_Def}).  Observe that in tensor network notation we have,
 \begin{align}
 \left(V^L_j(g)\otimes V^R_k(g)\right) U_{[j,k]}(g)|\psi\rangle &=~~ \vcenter{\hbox{\includegraphics[width=0.35\linewidth]{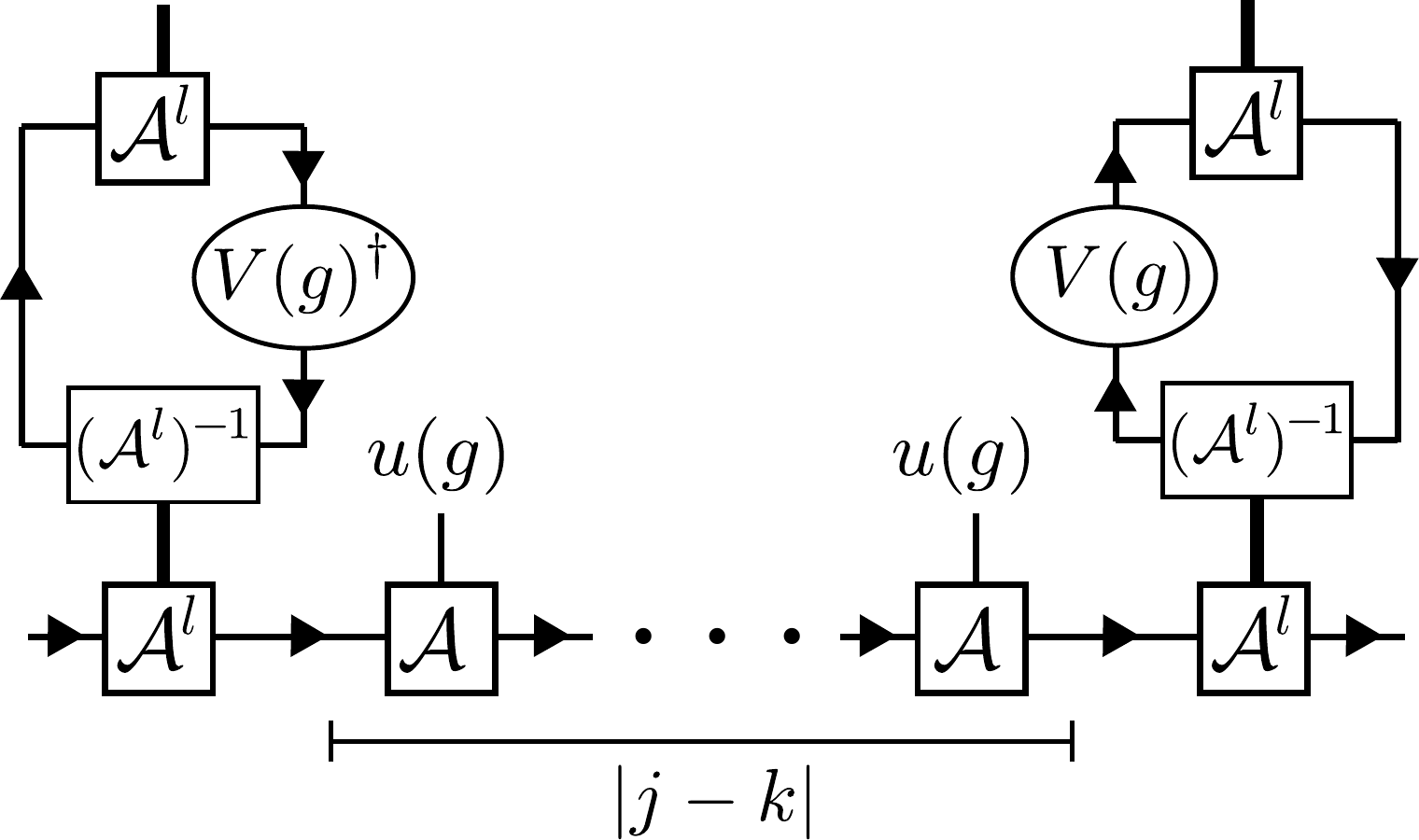}}} \label{TS_Der_1} \\
 &=~~ \vcenter{\hbox{\includegraphics[width=0.55\linewidth]{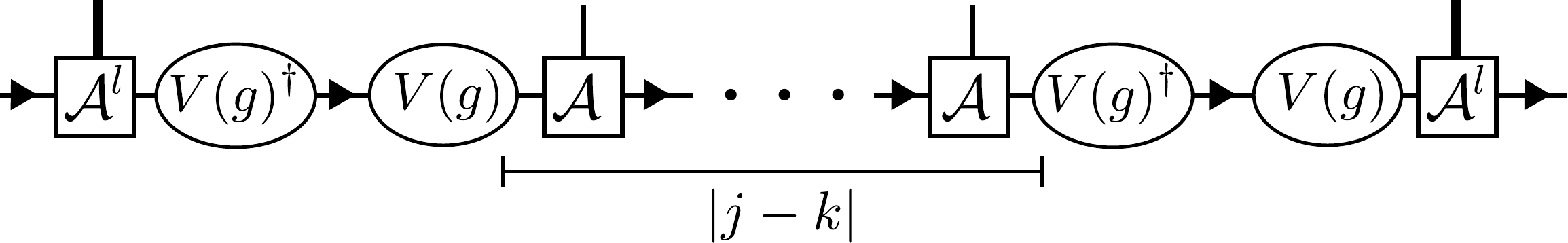}}} \label{TS_Der_2} \\
 &=~~ \vcenter{\hbox{\includegraphics[width=0.35\linewidth]{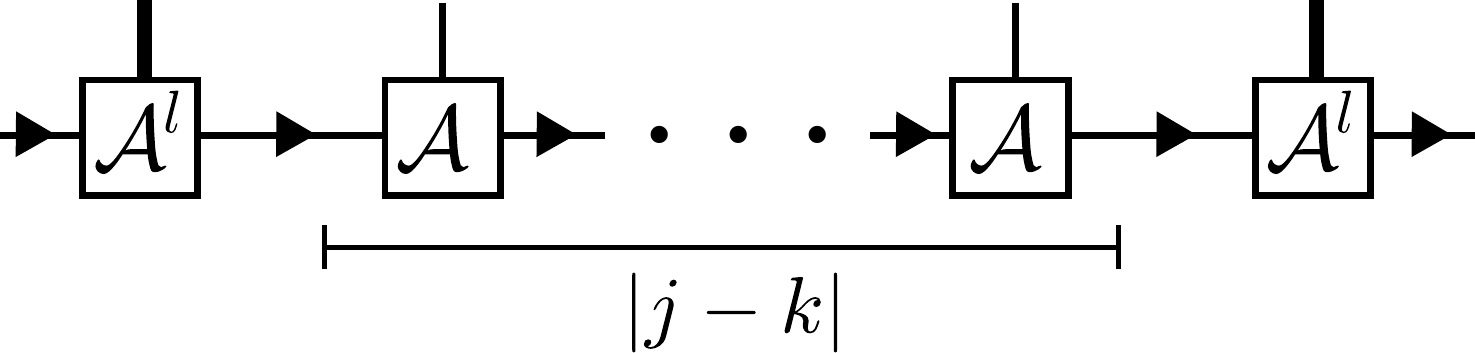}}} \label{TS_Der_3}\\
 &= |\psi\rangle.~~~~~\square \label{TS_Der_4}
 \end{align}
In Eq.~(\ref{TS_Der_1}) we have written the expression in tensor network notation using Eqs.~(\ref{VL_TNN}) and (\ref{VR_TNN}).  In Eq.~(\ref{TS_Der_2}) we utilized Eq.~({\ref{L_Inv_Def}}) and (\ref{MPS_Prep}) to push though $V^L(g)$, $V^R(g)$, and each $u(g)$ to the virtual space where each unitary $V(g)$ cancels with its inverse giving Eq.(\ref{TS_Der_3}), which is simply the state $|\psi\rangle$.

\item Next, we prove Eq.~(\ref{R_Proj}). This follows from a direct calculation.  In tensor network notation,
\begin{align}
V^R(g)V^R(h) = \vcenter{\hbox{\includegraphics[width=0.1\linewidth]{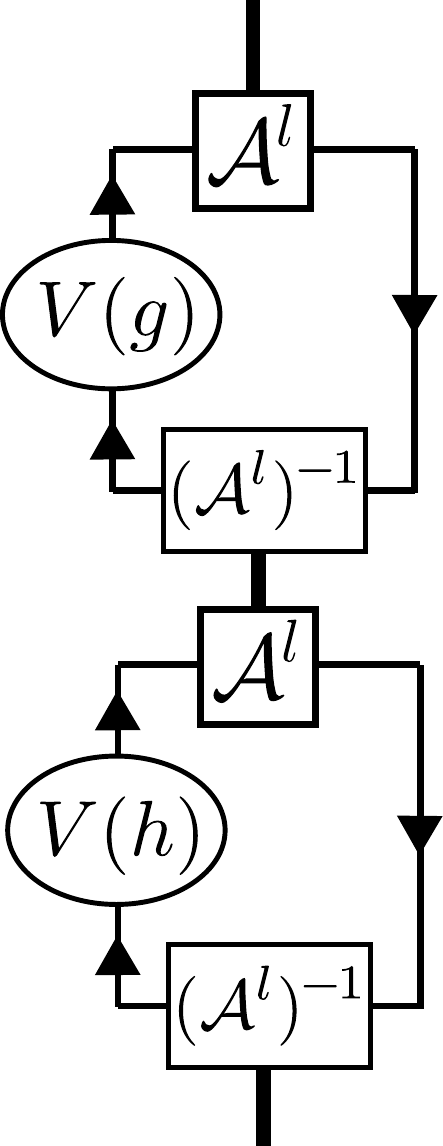}}} = \vcenter{\hbox{\includegraphics[width=0.1\linewidth]{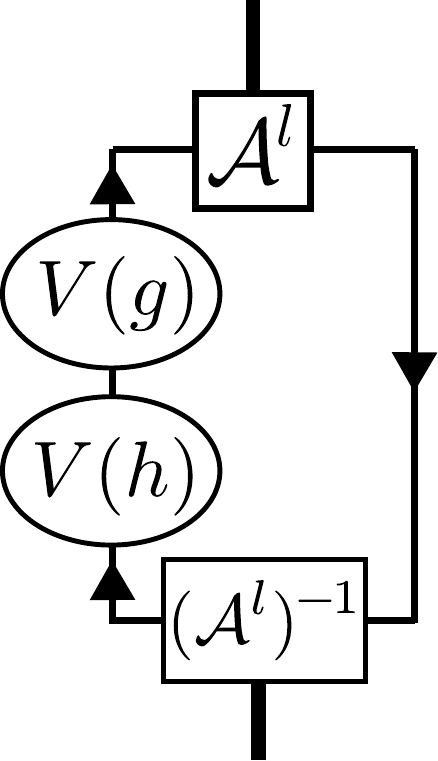}}} = \omega(g,h)\vcenter{\hbox{\includegraphics[width=0.1\linewidth]{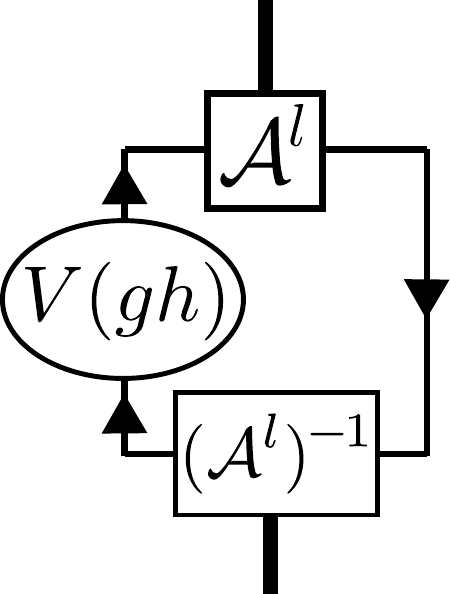}}} = \omega(g,h)V^R(gh).
\end{align}

\item Next, we prove Eq.~(\ref{L_Proj}).  This follows from a direct calculation.  In tensor network notation,
\begin{align}
V^L(h)V^L(g) &= \vcenter{\hbox{\includegraphics[width=0.1\linewidth]{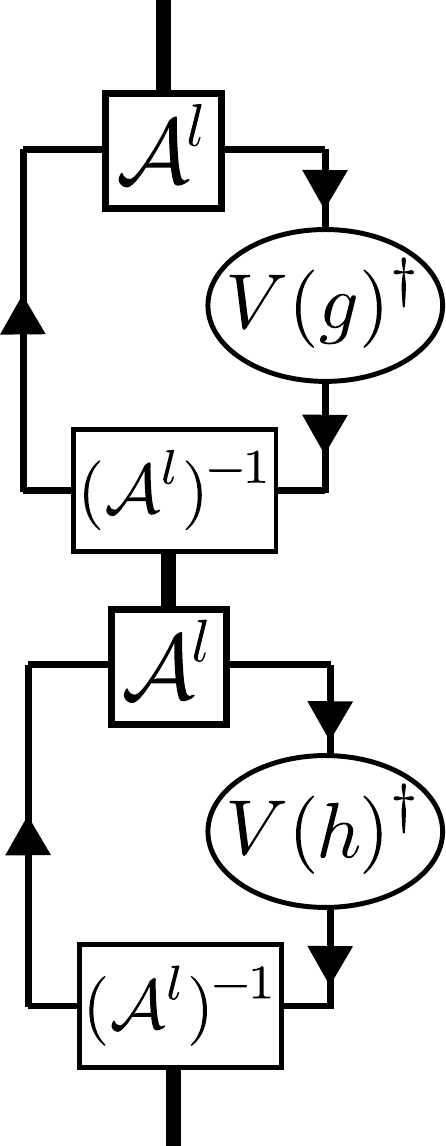}}} = \vcenter{\hbox{\includegraphics[width=0.1\linewidth]{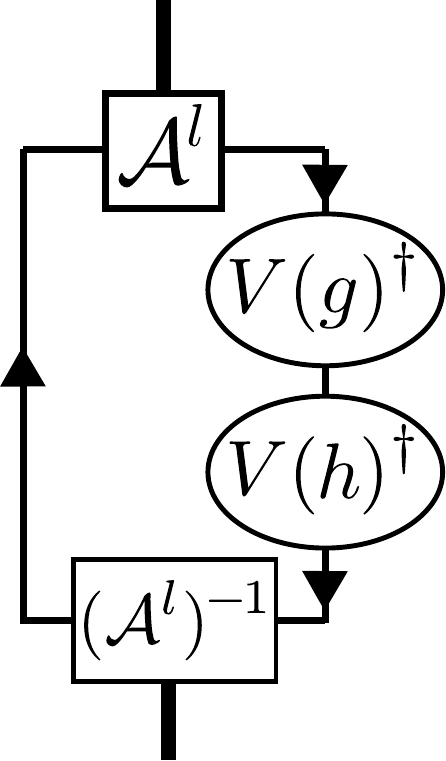}}} =\frac{1}{\omega(g,h)}\vcenter{\hbox{\includegraphics[width=0.1\linewidth]{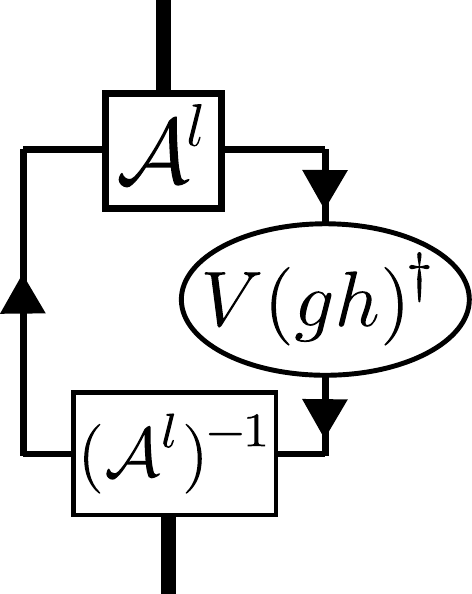}}} = \frac{1}{\omega(g,h)}V^L(gh).
\end{align}

\item Next, we prove Eq.~(\ref{RL_Com}).  This follows from a direct calculation.  In tensor network notation,
\begin{align}
V^R(g)V^L(h) &= \vcenter{\hbox{\includegraphics[width=0.13\linewidth]{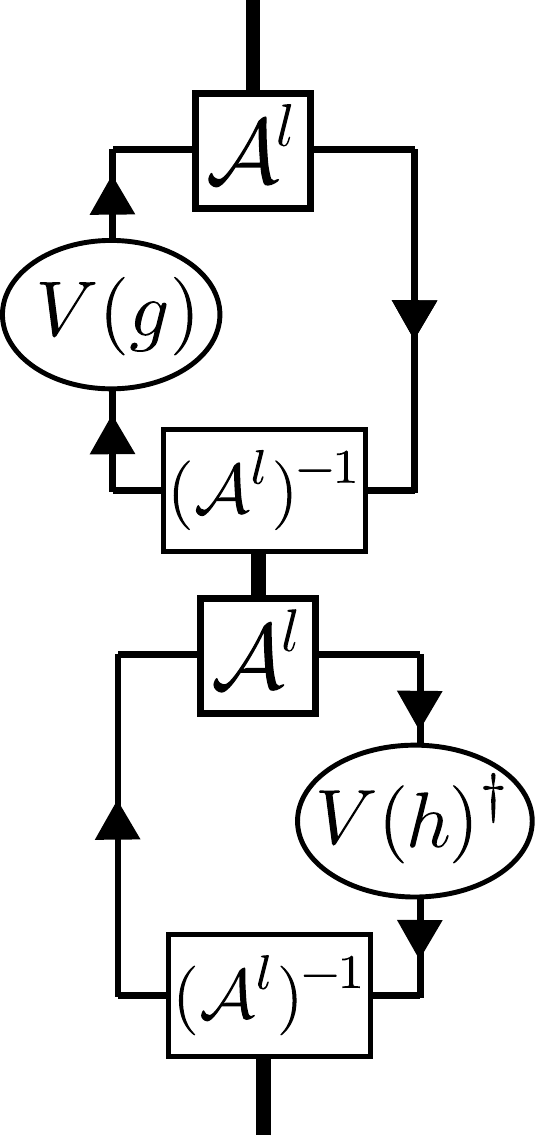}}} = \vcenter{\hbox{\includegraphics[width=0.13\linewidth]{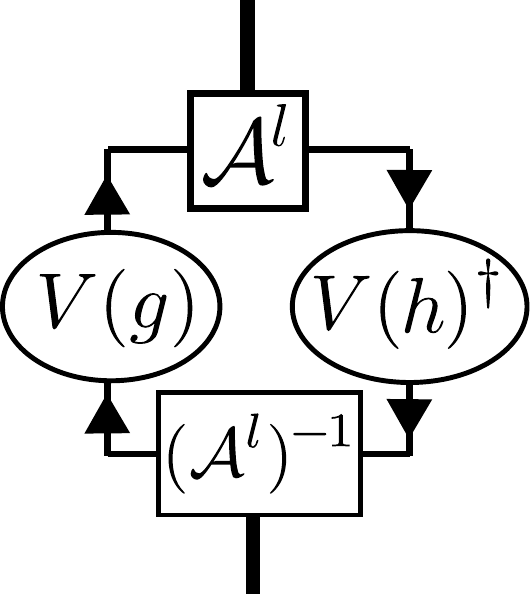}}} = \vcenter{\hbox{\includegraphics[width=0.13\linewidth]{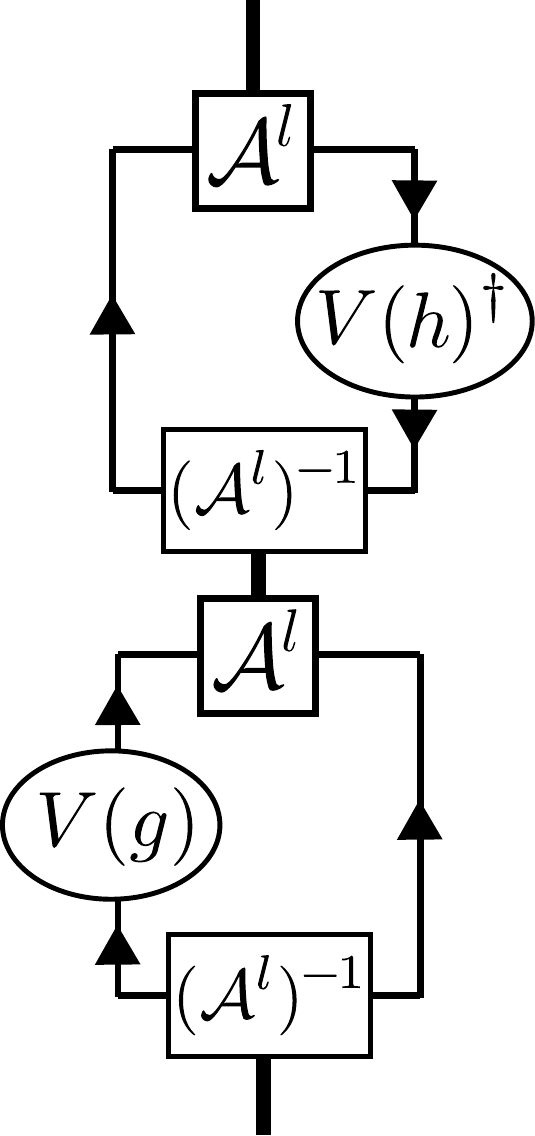}}} = V^L(h) V^R(g).
\end{align}

\item Next, we prove Eq.~(\ref{RL_Sym}).  This follows from a direct calculation.  In tensor network notation,
\begin{align}
V^R(g)V^L(g) &=  \vcenter{\hbox{\includegraphics[width=0.13\linewidth]{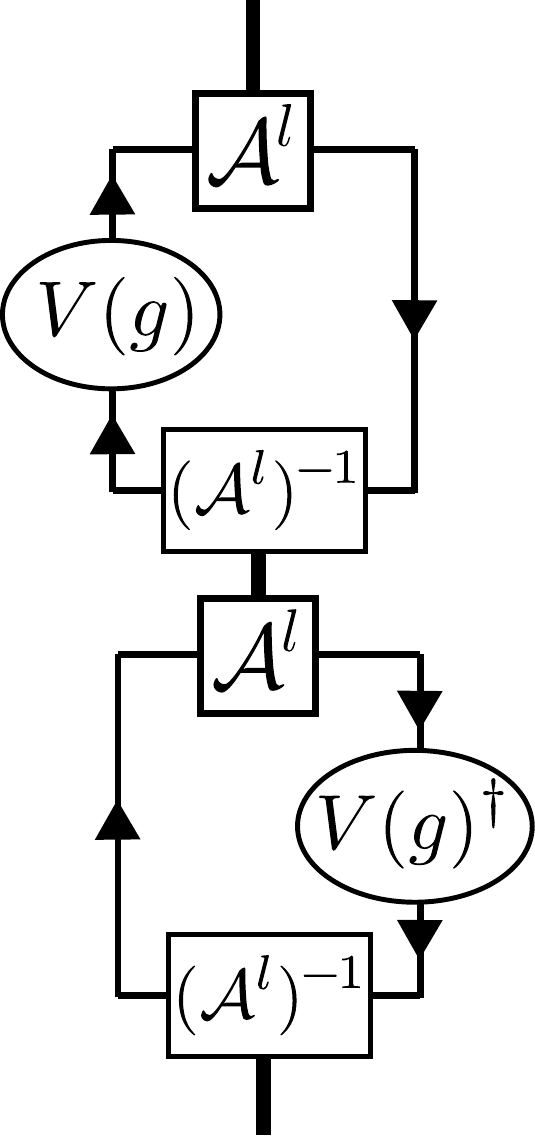}}} = \vcenter{\hbox{\includegraphics[width=0.13\linewidth]{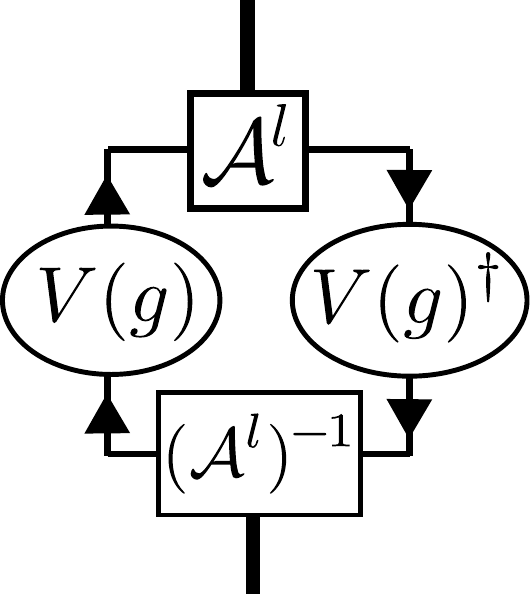}}} = \vcenter{\hbox{\includegraphics[width=0.085\linewidth]{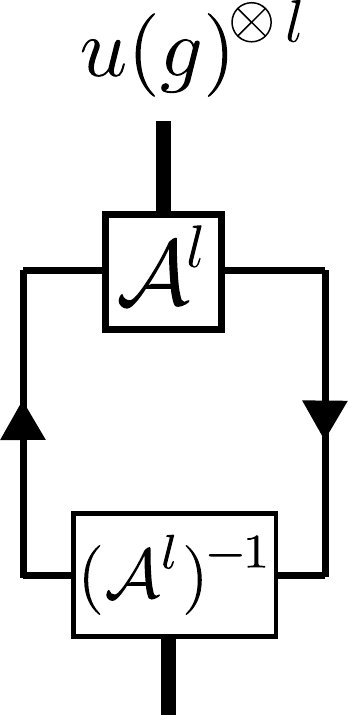}}}.
\end{align}
The left inverse also satisfies $\mathcal{A}^l(\mathcal{A}^l)^{-1} = \Pi_{\textrm{Range}(\mathcal{A})}$ where $\Pi_{\textrm{Range}(\mathcal{A})}$ is the projector onto the range of $\mathcal{A}$.  Thus,
\begin{align}
\vcenter{\hbox{\includegraphics[width=0.085\linewidth]{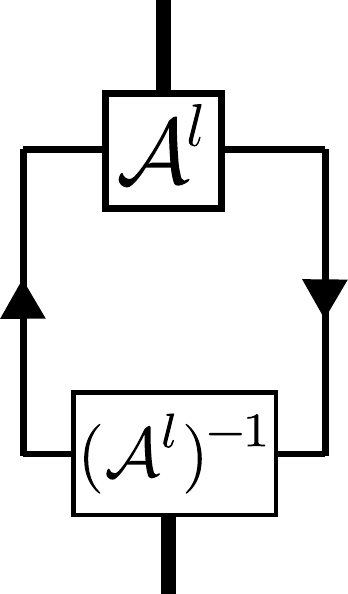}}} = \Pi_{\textrm{Range}(\mathcal{A})}.
\end{align}
Since $|\psi\rangle$ is the MPS generated by $\mathcal{A}$, we have $|\psi\rangle\in\textrm{Range}(\mathcal{A})$.  Hence $V^R(g)V^L(g) = u(g)^{\otimes l}$. $\square$

\item Finally, we prove Eq.~(\ref{Twist_Phase_Expectation}).  This follows from a direct calculation.  In tensor network notation,
\begin{align}
V_j^R(g)U(h)V_k^L(g) U_{[j,k]}(g) |\psi\rangle &= \vcenter{\hbox{\includegraphics[width=0.48\linewidth]{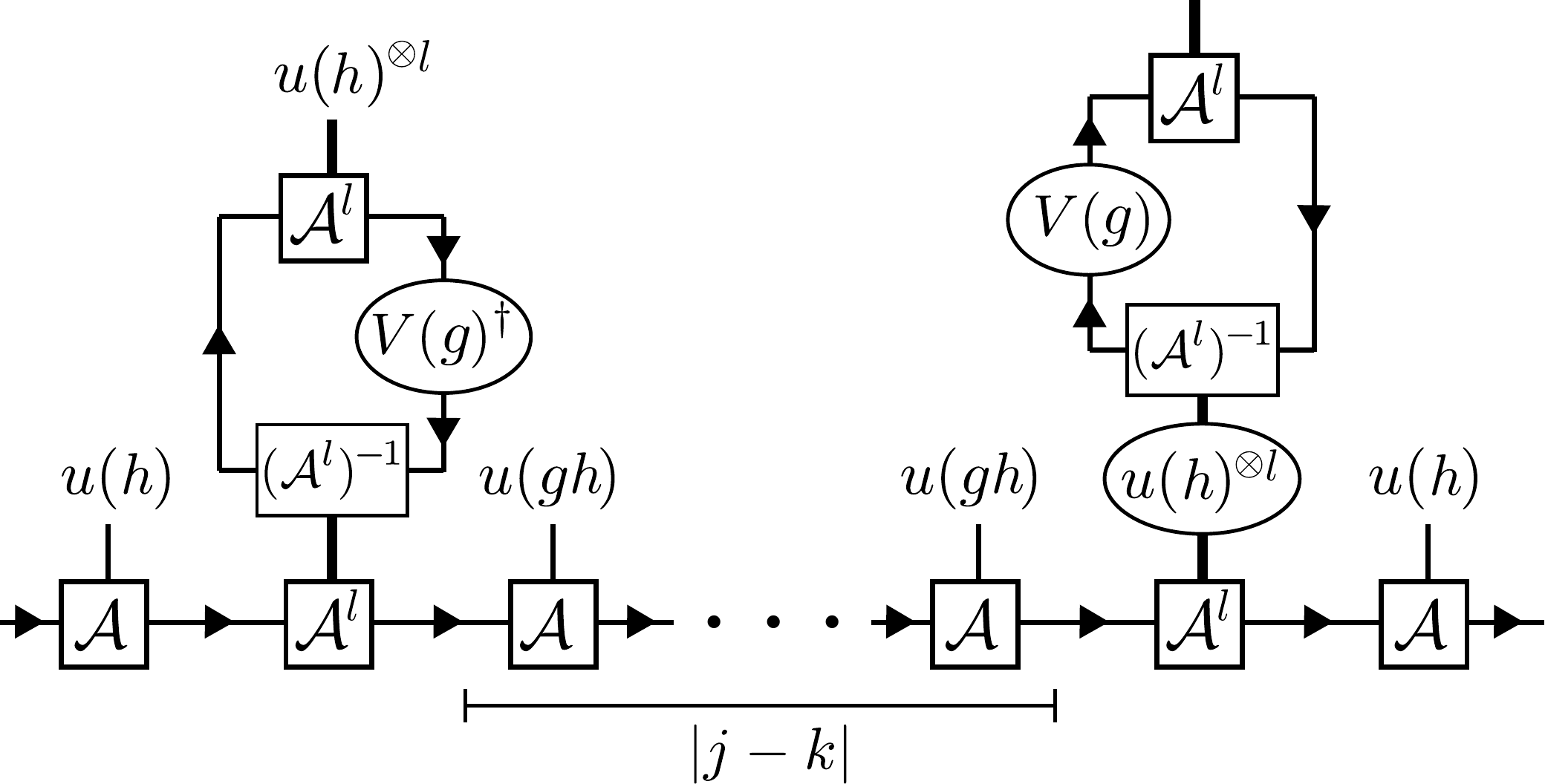}}} \label{Twist_Der_1} \\
&= \vcenter{\hbox{\includegraphics[width=0.55\linewidth]{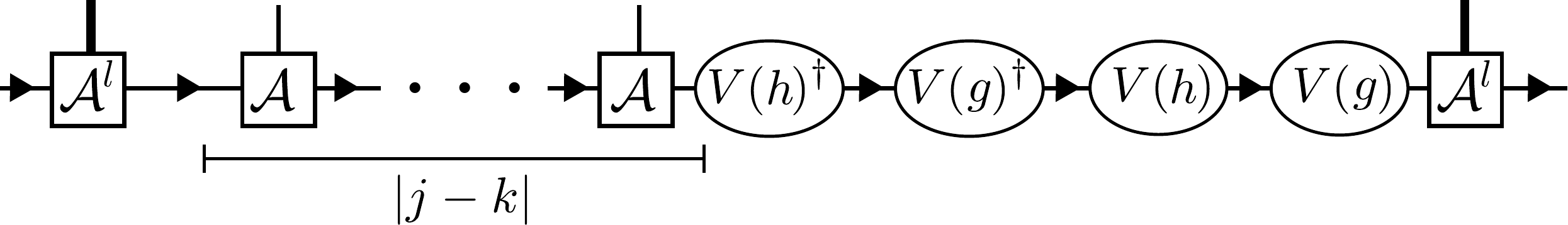}}} \label{Twist_Der_2} \\
&= \vcenter{\hbox{\includegraphics[width=0.47\linewidth]{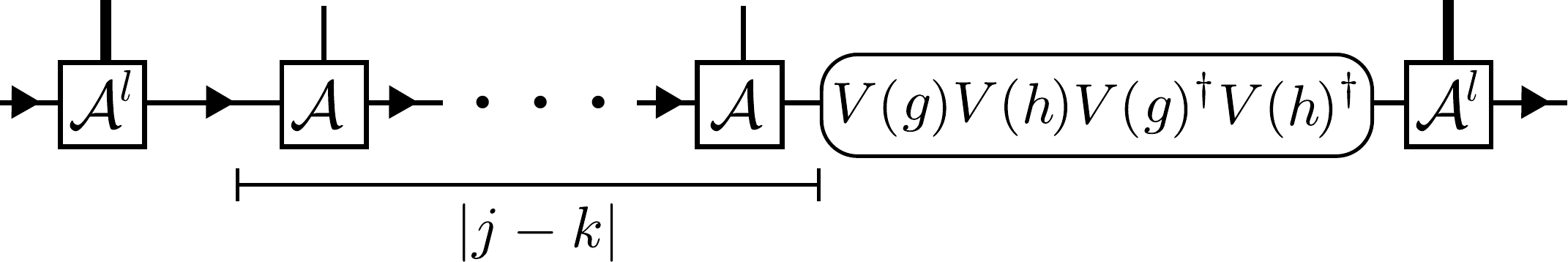}}} \label{Twist_Der_3} \\
&= \vcenter{\hbox{\includegraphics[width=0.4\linewidth]{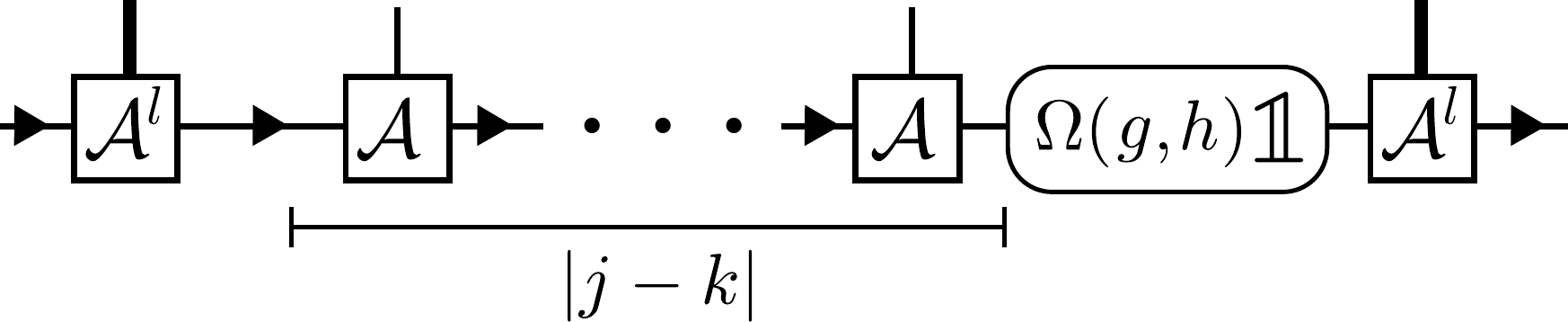}}} \label{Twist_Der_4} \\
&= \Omega(g,h) |\psi\rangle.
\end{align}
It then follows that $\langle \psi| T^{(g,h)}_{[j,k]}|\psi\rangle = \Omega(g,h)$.  $\square$
 \end{itemize}

\section{Fixed-point protocol at the AKLT point}
\label{SM3}

The AKLT state is the ground state of a 2-local spin-1 Hamiltonian of antiferromagnetic Heisenberg-type interactions
\begin{align}
H_{\mathrm{AKLT}} = \sum_{j=0}^{n-1} \mathbf{S}_j\cdot\mathbf{S}_{j+1} + \frac{1}{3}\left(\mathbf{S}_j\cdot\mathbf{S}_{j+1}\right)^2.
\end{align}
Here $\mathbf{S}_j = (S_j^x,S_j^y,S_j^z)$ where $S_j^\mu$ is the $\mu\in\{x,y,z\}$ spin component operator for the $j^{\textrm{th}}$ spin-1 particle.  All subscripts indexing sites are taken modulo $n$. Upon noting that each Hamiltonian term is simply a projector onto the spin-2 subspace of the two spin-1 Hilbert space---recall that $3\otimes3 = 1\oplus3\oplus5$---up to a rescaling by $2/3$, it is clear that the AKLT state only has support on the singlet and triplet (i.e. spin-0 and spin-1) subspaces.

The AKLT state has an MPS representation generated by the following three-index tensor,
\begin{align}
\mathcal{A}_{\textrm{AKLT}} = \frac{1}{\sqrt{3}} \sum_{\mu\in\{x,y,z\}} \sigma_\mu \otimes |\mu\rangle.
\end{align}
Here $\sigma_\mu$ denotes the Pauli matrices and $|\mu\rangle$ denotes the spin-1 cartesian basis defined by $S^\mu|\mu\rangle = 0$ for each ${\mu\in\{x,y,z\}}$.  In the eigenbasis of $S^z$ (i.e. $\{|-1\rangle,|0\rangle,|+1\rangle\}$) the cartesian basis vectors are
\begin{subequations}
\begin{align}
|x\rangle &= \frac{-|+1\rangle + |-1\rangle}{\sqrt{2}} \\
|y\rangle &= i\frac{|+1\rangle + |-1\rangle}{\sqrt{2}} \\
|z\rangle &= |0\rangle.
\end{align}
\end{subequations}
The AKLT state has a $\mathbb{Z}_2\times\mathbb{Z}_2$ symmetry with onsite representation given by $u(\mu) = \exp(i\pi S^\mu)$ for each $\mu\in\{x,y,z\}$.  It is easily checked that the action of $u(x)$, $u(y)$, or $u(z)$ on the physical index of $\mathcal{A}_\mathrm{AKLT}$ is equivalent to conjugation by $\sigma_x$, $\sigma_y$, or $\sigma_z$ on the virtual indices, respectively.  The Pauli matrices form a projective representation of $\mathbb{Z}_2\times\mathbb{Z}_2$ and thus the AKLT state has 1D $\mathbb{Z}_2\times\mathbb{Z}_2$ SPTO.

This MPS has transfer matrix $\mathcal{E}_{\mathcal{A}_{\textrm{AKLT}}}=(\sigma_x\otimes\sigma_x - \sigma_y\otimes\sigma_y + \sigma_z\otimes\sigma_z)/3$, which is diagonal in the Bell-basis $\{|\phi^+\rangle, |\phi^-\rangle, |\psi^+\rangle, |\psi^-\rangle\}$ giving $\mathcal{E}_{\mathcal{A}_{\textrm{AKLT}}} = \textrm{diag}(1,1/3,1/3,1/3)$.
Since the correlation length for this MPS is $\xi = 1/\ln(3)$, it resides outside the fixed-point of the $\mathbb{Z}_2\times\mathbb{Z}_2$ SPTO phase.
This MPS is injective, with injectivity length 2.  The two site MPS tensor is given by,
\begin{align}
\mathcal{A}_{\textrm{AKLT}}^2 = \frac{1}{\sqrt{3}}\mathbbm{1}\otimes|\tilde{e}\rangle + \sqrt{\frac{2}{9}}\sum_{\mu\in\{x,y,z\}} \sigma_\mu \otimes|\tilde{\mu}\rangle.
\label{AKLT_Two_Site}
\end{align}
Where $|\tilde{e}\rangle = \frac{1}{\sqrt{3}}\sum_{\nu\in\{x,y,z\}} |\nu\nu\rangle$ and $|\tilde{\mu}\rangle = \frac{i}{\sqrt{2}}\sum_{\nu,\gamma\in\{x,y,z\}}\epsilon_{\mu\nu\gamma} |\nu\gamma\rangle$.  Notice that $|\tilde{e}\rangle$ is the singlet for two spin-1 particles.  Furthermore, $|\tilde{\mu}\rangle$ for $\mu\in\{x,y,z\}$ span the two spin-1 triplet.  Furthermore, these form the cartesian basis in the sense that $(S_1^\mu + S_2^\mu)|\tilde{\mu}\rangle = 0$.  It is easily checked that the two site transfer matrix is simply the square of the single site transfer matrix, $\mathcal{E}_{\mathcal{A}_{\textrm{AKLT}}^2} = \mathcal{E}_{\mathcal{A}_{\textrm{AKLT}}}^2 = \textrm{diag}(1,1/9,1/9,1/9)$.

The fixed-point MPS approached by the AKLT state under RG flows has a transfer matrix $\lim_{M\rightarrow\infty}\mathcal{E}_{\mathcal{A}_{\textrm{AKLT}}}^M =  \textrm{diag}(1,0,0,0)$.  The fixed-point MPS with this transfer matrix is
\begin{align}
\mathcal{A}_\textrm{AKLT, Fix}^2 = \frac{1}{2}\mathbbm{1}\otimes|\tilde{e}\rangle + \frac{1}{2}\sum_{\mu\in\{x,y,z\}} \sigma_\mu \otimes|\tilde{\mu}\rangle.
\end{align}
Applying the local isometry $\Pi = |++\rangle\langle\tilde{e}| + |-+\rangle\langle\tilde{z}| + |+-\rangle\langle\tilde{x}| - i|--\rangle\langle\tilde{y}|$ to the physical degrees of freedom gives an MPS tensor that generates the 1D cluster state.  We remark that this mapping replaces each pair of spin-1 particles with a pair of qubits.  The fixed-point boundary operators for $\mu\in\{x,y,z\}$ are thus equivalent to those given for the 1D cluster state in the main text under the aforementioned isometry. These are given as
\begin{subequations}
\begin{align}
{V}^L(\mu) &= |\tilde{\mu}\rangle\langle\tilde{e}| +  |\tilde{e}\rangle\langle\tilde{\mu}| - i\sum_{\nu\gamma}\epsilon_{\mu\nu\gamma}|\tilde{\nu}\rangle\langle\tilde{\gamma}| \label{VL_AKLT}\\
{V}^R(\mu) &= |\tilde{\mu}\rangle\langle\tilde{e}| + |\tilde{e}\rangle\langle\tilde{\mu}| + i\sum_{\nu\gamma}\epsilon_{\mu\nu\gamma} |\tilde{\nu}\rangle\langle\tilde{\gamma}\label{VR_AKLT}|,
\end{align}
\end{subequations}
which are unitary and Hermitian.  Furthermore, $V^R(e)=V^L(e)=\mathbbm{1}$.
Explicitly for $\mu = z$ we have,
\begin{subequations}
\begin{align}
{V}^L(z) &=  |\tilde{z}\rangle\langle\tilde{e}| +  |\tilde{e}\rangle\langle\tilde{z}| - i |\tilde{x}\rangle\langle\tilde{y}| +i |\tilde{y}\rangle\langle\tilde{x}| \label{VLz}\\
{V}^R(z) &=  |\tilde{z}\rangle\langle\tilde{e}| +  |\tilde{e}\rangle\langle\tilde{z}| + i |\tilde{x}\rangle\langle\tilde{y}| - i |\tilde{y}\rangle\langle\tilde{x}|\label{VRz}.
\end{align}
\end{subequations}
It can be checked that the above fixed-point boundary operators can be expressed in terms of spin-1 observables as
\begin{subequations}
\begin{align}
V^L(\mu) &= \frac{1}{\sqrt{24}}\left\{\frac{(S^2-4)}{2},S_1^\mu - S_2^\mu \right\} + \frac{1}{2}\left\{\frac{S^2(6-S^2)}{8}, S_1^\mu + S_2^\mu\right\} \label{VL_AKLT_spin}\\
V^R(\mu) &= \frac{1}{\sqrt{24}}\left\{\frac{(S^2-4)}{2},S_1^\mu - S_2^\mu \right\} - \frac{1}{2}\left\{\frac{S^2(6-S^2)}{8}, S_1^\mu + S_2^\mu\right\} \label{VR_AKLT_spin}.
\end{align}
\end{subequations}
Here $S^2 = (\mathbf{S}_1 + \mathbf{S}_2)^2$ and $\{\cdot,\cdot\}$ denotes the anticommutator.  We remark that for MPS calculations it is convenient to work with Eqs.~(\ref{VL_AKLT})--(\ref{VR_AKLT}); however, Eqs.~(\ref{VL_AKLT_spin})--(\ref{VR_AKLT_spin}) inform how the boundary operators can be measured in an experiment.

Expectation values of strings of local operators can be conveniently calculated in the MPS formalism via transfer matrices.  For an MPS generated by tensor $\mathcal{A} = \sum_{j=0}^{d-1} A^{(j)} \otimes|j\rangle$ and for any single-site operator $O\in L(\mathbb{C}^d)$, the transfer matrix $\mathcal{E}_O$ is defined as,
\begin{align}
\mathcal{E}_O = \sum_{j,k=0}^{d-1} \langle j|O|k\rangle A^{(k)}\otimes{A^{(j)}}^*.
\end{align}
The expected value of $O$ in the state $|\psi\rangle$, consisting of $N$ sites, can be written,
\begin{align}
\langle\psi|O|\psi\rangle = \textrm{Tr}(\mathcal{E}_\mathbbm{1}^{N-1}\mathcal{E}_O)/ \textrm{Tr}(\mathcal{E}_\mathbbm{1}^{N}).
\end{align}

To compute the expected value of the string order parameter of Eq.~(\ref{Trunc_Sym}) at the AKLT point, we must analyze the transfer matrices $\mathcal{E}_{u(g)}$, $\mathcal{E}_{V^L(g)}$, and $\mathcal{E}_{V^R(g)}$.  Notice that $\mathcal{E}_{V^L(g)}$ and $\mathcal{E}_{V^R(g)}$ are obtained from the two-site tensor in Eq.~(\ref{AKLT_Two_Site}).  Taking $g=z$, these transfer matrices are written,
\begin{subequations}
\begin{align}
\mathcal{E}_{u(z)} &=
\begin{pmatrix}
-\frac{1}{3} & 0 & 0 & 0 \\
0 & -\frac{1}{3} & 0 & 0 \\
0 & 0 & -\frac{1}{3} & 0 \\
0 & 0 & 0 & 1
\end{pmatrix} ,\\
\mathcal{E}_{{V}^L(z)} &= 
\begin{pmatrix}
0 & 0 & 0 & \frac{2}{3}\left( \sqrt{\frac{2}{3}}-\frac{2}{3}\right) \\
0 & 0 & 0 & 0 \\
0 & 0 & 0 & 0 \\
\frac{2}{3}\left( \sqrt{\frac{2}{3}}+\frac{2}{3}\right) & 0 & 0 & 0
\end{pmatrix} ,
\\
\mathcal{E}_{{V}^R(z)} &= 
\begin{pmatrix}
0 & 0 & 0 & \frac{2}{3}\left( \sqrt{\frac{2}{3}}+\frac{2}{3}\right) \\
0 & 0 & 0 & 0 \\
0 & 0 & 0 & 0 \\
\frac{2}{3}\left( \sqrt{\frac{2}{3}}-\frac{2}{3}\right) & 0 & 0 & 0
\end{pmatrix}.
\end{align}
\end{subequations}
The expected value string order parameter of Eq.~(\ref{Trunc_Sym}) can then be written as,
\begin{align}
\langle S_{[j,k]}(z)\rangle_{\textrm{AKLT}} &= \textrm{Tr}\left(\mathcal{E}_\mathbbm{1}^{N-|j-k|-4}\mathcal{E}_{V^R(z)}\mathcal{E}_{u(z)}^{|j-k|} \mathcal{E}_{V^L(z)}\right)  \nonumber \\
&= \frac{4}{9}\left( \sqrt{\frac{2}{3}} + \frac{2}{3} \right)^2 + \frac{4}{9}\left( \sqrt{\frac{2}{3}} - \frac{2}{3} \right)^2\left(-\frac{1}{3} \right)^{N-4}.
\end{align}
The MPS tensor is symmetric with respect to permutations of $\{x,y,z\}$ and thus the value is the same for any nontrivial group element $g\in\{x,y,z\}$.  Thus, the AKLT state on an even number of sites, $N>4$, has $\frac{4}{9}\left( \sqrt{\frac{2}{3}} + \frac{2}{3} \right)^2 < \langle S_{[j,k]}(g)\rangle_{\textrm{AKLT}} < \frac{80}{81}$.  Since the AKLT state possesses the $\mathbb{Z}_2\times\mathbb{Z}_2$ symmetry, this calculation carries over for the twist operator since by Eq.~(\ref{RL_Sym}), $\langle T_{[j,k]}^{(g,h)}\rangle_{\textrm{AKLT}} = \Omega(g,h) \langle S_{[j,k]}(g)\rangle_{\textrm{AKLT}} $.  Five of eight instances of the triangle game require measurement of the symmetry---which is perfect at the AKLT point---and the other three require measurement of the twist operator.  Therefore, the average success probability for the SPTO strategy applied at the AKLT point is,
\begin{align}
\textrm{pr}(\textrm{win}|\textrm{AKLT}) &= \frac{13}{16} + \frac{1}{12}\left(\sqrt{\frac{2}{3}}+\frac{2}{3}\right)^2 + \frac{1}{12} \left(\sqrt{\frac{2}{3}}-\frac{2}{3}\right)^2\left(-\frac{1}{3}\right)^{N-4} \nonumber \\
& \stackrel{N\rightarrow\infty}{\longrightarrow} \frac{13}{16} + \frac{1}{12}\left(\sqrt{\frac{2}{3}}+\frac{2}{3}\right)^2 \approx 0.996.
\end{align}

\bibliography{ref.bib}